\documentclass[twocolumn]{aastex62}
\submitjournal{ApJ}

\shorttitle{FAUST III. Misaligned rotations of the envelope, outflow, and disks}
\shortauthors{Ohashi et al.}
\graphicspath{{./}{figures/}}

\begin{document}

\title{FAUST III. Misaligned rotations of the envelope, outflow, and disks  in the multiple protostellar system of VLA 1623$-$2417}

\author[0000-0002-9661-7958]{Satoshi Ohashi}
\affil{RIKEN Cluster for Pioneering Research, 2-1, Hirosawa, Wako-shi, Saitama 351-0198, Japan}
\email{satoshi.ohashi@riken.jp}

\author{Claudio Codella}
\affil{INAF, Osservatorio Astrofisico di Arcetri, Largo E. Fermi 5, I-50125, Firenze, Italy}
\affil{Univ. Grenoble Alpes, CNRS, IPAG, 38000 Grenoble, France}

\author{Nami Sakai}
\affiliation{RIKEN Cluster for Pioneering Research, 2-1, Hirosawa, Wako-shi, Saitama 351-0198, Japan}

\author{Claire J. Chandler}
\affiliation{National Radio Astronomy Observatory, PO Box O, Socorro, NM 87801, USA}

\author{Cecilia Ceccarelli}
\affiliation{Univ. Grenoble Alpes, CNRS, IPAG, 38000 Grenoble, France}

\author{Felipe Alves}
\affiliation{Center for Astrochemical Studies, Max-Planck-Institut f\"{u}r extraterrestrische Physik (MPE), Gie$\beta$enbachstr. 1, D-85741 Garching, Germany}

\author{Davide Fedele}
\affiliation{INAF, Osservatorio Astrofisico di Arcetri, Largo E. Fermi 5, I-50125, Firenze, Italy}

\author{Tomoyuki Hanawa}
\affiliation{Center for Frontier Science, Chiba University, 1-33 Yayoi-cho, Inage-ku, Chiba 263-8522, Japan}

\author{Aurora Dur\'{a}n}
\affiliation{Instituto de Radioastronom\'{i}a y Astrof\'{i}sica , Universidad Nacional Aut\'{o}noma de M\'{e}xico, A.P. 3-72 (Xangari), 8701, Morelia, Mexico}
\author{C\'{e}cile Favre}
\affiliation{Univ. Grenoble Alpes, CNRS, IPAG, 38000 Grenoble, France}

\author{Ana L\'{o}pez-Sepulcre}
\affiliation{Univ. Grenoble Alpes, CNRS, IPAG, 38000 Grenoble, France}
\affiliation{Institut de Radioastronomie Millim\'{e}trique, 38406 Saint-Martin d'H$\grave{e}$res, France}

\author{Laurent Loinard}
\affiliation{Instituto de Radioastronom\'{i}a y Astrof\'{i}sica , Universidad Nacional Aut\'{o}noma de M\'{e}xico, A.P. 3-72 (Xangari), 8701, Morelia, Mexico}
\affiliation{Instituto de Astronom\'{i}a, Universidad Nacional Aut\'{o}noma de M\'{e}xico, Ciudad Universitaria, A.P. 70-264, Cuidad de M\'{e}xico 04510, Mexico}

\author{Seyma Mercimek}
\affiliation{INAF, Osservatorio Astrofisico di Arcetri, Largo E. Fermi 5, I-50125, Firenze, Italy}
\affiliation{Universit$\grave{a}$ degli Studi di Firenze, Dipartimento di Fisica e Astronomia, via G. Sansone 1, 50019 Sesto Fiorentino, Italy}

\author{Nadia M. Murillo}
\affiliation{RIKEN Cluster for Pioneering Research, 2-1, Hirosawa, Wako-shi, Saitama 351-0198, Japan}

\author{Linda Podio}
\affiliation{INAF, Osservatorio Astrofisico di Arcetri, Largo E. Fermi 5, I-50125, Firenze, Italy}

\author{Yichen Zhang}
\affiliation{RIKEN Cluster for Pioneering Research, 2-1, Hirosawa, Wako-shi, Saitama 351-0198, Japan}

\author{Yuri Aikawa}
\affiliation{Department of Astronomy, The University of Tokyo, 7-3-1 Hongo, Bunkyo-ku, Tokyo 113-0033, Japan}

\author{Nadia Balucani}
\affiliation{Department of Chemistry, Biology, and Biotechnology, The University of Perugia, Via Elce di Sotto 8, 06123 Perugia, Italy}
\author{Eleonora Bianchi}
\affiliation{Univ. Grenoble Alpes, CNRS, IPAG, 38000 Grenoble, France}
\author{Mathilde Bouvier}
\affiliation{Univ. Grenoble Alpes, CNRS, IPAG, 38000 Grenoble, France}

\author{Gemma Busquet}
\affil{Univ. Grenoble Alpes, CNRS, IPAG, 38000 Grenoble, France}

\author{Paola Caselli}
\affiliation{Center for Astrochemical Studies, Max-Planck-Institut f\"{u}r extraterrestrische Physik (MPE), Gie$\beta$enbachstr. 1, D-85741 Garching, Germany}

\author{Emmanuel Caux}
\affiliation{IRAP, Universit\'{e} de Toulouse, CNRS, CNES, UPS, Toulouse, France}
\author{Steven Charnley}
\affiliation{Astrochemistry Laboratory, Code 691, NASA Goddard Space Flight Center, 8800 Greenbelt Road, Greenbelt, MD 20771, USA}
\author{Spandan Choudhury}
\affiliation{Center for Astrochemical Studies, Max-Planck-Institut f\"{u}r extraterrestrische Physik (MPE), Gie$\beta$enbachstr. 1, D-85741 Garching, Germany}

\author{Nicolas Cuello}
\affiliation{Univ. Grenoble Alpes, CNRS, IPAG, 38000 Grenoble, France}

\author{Marta De Simone}
\affiliation{Univ. Grenoble Alpes, CNRS, IPAG, 38000 Grenoble, France}
\author{Francois Dulieu}
\affiliation{CY Cergy Paris Universit\'{e}, Sorbonne Universit\'{e}, Observatoire de Paris, PSL University, CNRS, LERMA, F-95000, Cergy, France}

\author{Lucy Evans}
\affiliation{IRAP, Universit\'{e} de Toulouse, CNRS, CNES, UPS, Toulouse, France}
\affiliation{INAF, Osservatorio Astrofisico di Arcetri, Largo E. Fermi 5, I-50125, Firenze, Italy}

\author{Siyi Feng}
\affiliation{CAS Key Laboratory of FAST, National Astronomical Observatory of China, Datun Road 20, Chaoyang, Beijing, 100012, P. R. China}
\affiliation{National Astronomical Observatory of Japan, Osawa 2-21-1, Mitaka-shi, Tokyo 181-8588, Japan}
\affiliation{Institute of Astronomy and Astrophysics, Academia Sinica, 11F of Astronomy-Mathematics Building, AS/NTU No.1, Sec. 4, Roosevelt Rd., Taipei 10617, Taiwan}
\author{Francesco Fontani}
\affiliation{INAF, Osservatorio Astrofisico di Arcetri, Largo E. Fermi 5, I-50125, Firenze, Italy}
\affiliation{Center for Astrochemical Studies, Max-Planck-Institut f\"{u}r extraterrestrische Physik (MPE), Gie$\beta$enbachstr. 1, D-85741 Garching, Germany}
\author{Logan Francis}
\affiliation{NRC Herzberg Astronomy and Astrophysics, 5071 West Saanich Road, Victoria, BC, V9E 2E7, Canada}
 \affiliation{Department of Physics and Astronomy, University of Victoria, Victoria, BC, V8P 5C2, Canada}
\author{Tetsuya Hama}
\affiliation{Komaba Institute for Science, The University of Tokyo, 3-8-1 Komaba, Meguro, Tokyo 153-8902, Japan}
\affiliation{Department of Basic Science, The University of Tokyo, 3-8-1 Komaba, Meguro, Tokyo 153-8902, Japan}

\author{Eric Herbst}
\affiliation{Department of Chemistry, University of Virginia, McCormick Road, PO Box 400319, Charlottesville, VA 22904, USA}

\author{Shingo Hirano}
\affiliation{Department of Astronomy, The University of Tokyo, 7-3-1 Hongo, Bunkyo-ku, Tokyo 113-0033, Japan}

\author{Tomoya Hirota}
\affiliation{National Astronomical Observatory of Japan, Osawa 2-21-1, Mitaka-shi, Tokyo 181-8588, Japan}
\author{Muneaki Imai}
\affiliation{Department of Physics, The University of Tokyo, 7-3-1, Hongo, Bunkyo-ku, Tokyo 113-0033, Japan}
\author{Andrea Isella}
\affiliation{Department of Physics and Astronomy, Rice University, 6100 Main Street, MS-108, Houston, TX 77005, USA}
\author{Izaskun J\'{i}menez-Serra}
\affiliation{Centro de Astrobiolog\'{\i}a (CSIC-INTA), Ctra. de Torrej\'on a Ajalvir, km 4, 28850, Torrej\'on de Ardoz, Spain}
\author{Doug Johnstone}
\affiliation{NRC Herzberg Astronomy and Astrophysics, 5071 West Saanich Road, Victoria, BC, V9E 2E7, Canada}
\affiliation{Department of Physics and Astronomy, University of Victoria, Victoria, BC, V8P 5C2, Canada}
\author{Claudine Kahane}
\affiliation{Univ. Grenoble Alpes, CNRS, IPAG, 38000 Grenoble, France}

\author{Romane Le Gal}
\affiliation{IRAP, Université de Toulouse, CNRS, UPS, CNES, F-31400 Toulouse, France}

\author{Bertrand Lefloch}
\affiliation{Univ. Grenoble Alpes, CNRS, IPAG, 38000 Grenoble, France}

\author{Luke T. Maud}
\affiliation{European Southern Observatory, Karl-Schwarzschild Str. 2, 85748 Garching bei M\"{u}nchen, Germany}
\author{Maria Jose Maureira}
\affiliation{Center for Astrochemical Studies, Max-Planck-Institut f\"{u}r extraterrestrische Physik (MPE), Gie$\beta$enbachstr. 1, D-85741 Garching, Germany}
\author{Francois Menard}
\affiliation{Univ. Grenoble Alpes, CNRS, IPAG, 38000 Grenoble, France}

\author{Anna Miotello}
\affiliation{European Southern Observatory, Karl-Schwarzschild Str. 2, 85748 Garching bei M\"{u}nchen, Germany}
\author{George Moellenbrock}
\affiliation{National Radio Astronomy Observatory, PO Box O, Socorro, NM 87801, USA}
\author{Shoji Mori}
\affiliation{Department of Astronomy, The University of Tokyo, 7-3-1 Hongo, Bunkyo-ku, Tokyo 113-0033, Japan}

\author{Riouhei Nakatani}
\affiliation{RIKEN Cluster for Pioneering Research, 2-1, Hirosawa, Wako-shi, Saitama 351-0198, Japan}
\author{Hideko Nomura}
\affiliation{National Astronomical Observatory of Japan, Osawa 2-21-1, Mitaka-shi, Tokyo 181-8588, Japan}
\author{Yasuhiro Oba}
\affiliation{Institute of Low Temperature Science, Hokkaido University, N19W8, Kita-ku, Sapporo, Hokkaido 060-0819, Japan}
\author{Ross O'Donoghue}
\affiliation{Department of Physics and Astronomy, University College London, Gower Street, London, WC1E 6BT, UK}
\author{Yuki Okoda}
\affiliation{Department of Physics, The University of Tokyo, 7-3-1, Hongo, Bunkyo-ku, Tokyo 113-0033, Japan}
\author{Juan Ospina-Zamudio}
\affiliation{Univ. Grenoble Alpes, CNRS, IPAG, 38000 Grenoble, France}
\author{Yoko Oya}
\affiliation{Department of Physics, The University of Tokyo, 7-3-1, Hongo, Bunkyo-ku, Tokyo 113-0033, Japan}
\affiliation{Research Center for the Early Universe, The University of Tokyo, 7-3-1, Hongo, Bunkyo-ku, Tokyo 113-0033, Japan}

\author{Jaime Pineda}
\affiliation{Center for Astrochemical Studies, Max-Planck-Institut f\"{u}r extraterrestrische Physik (MPE), Gie$\beta$enbachstr. 1, D-85741 Garching, Germany}

\author{Albert Rimola}
\affiliation{Departament de Qu\'{i}mica, Universitat Aut$\grave{o}$noma de Barcelona, 08193 Bellaterra, Spain}
\author{Takeshi Sakai}
\affiliation{Graduate School of Informatics and Engineering, The University of Electro-Communications, Chofu, Tokyo 182-8585, Japan}
\author{Dominique Segura-Cox}
\affiliation{Center for Astrochemical Studies, Max-Planck-Institut f\"{u}r extraterrestrische Physik (MPE), Gie$\beta$enbachstr. 1, D-85741 Garching, Germany}
\author{Yancy Shirley}
\affiliation{Steward Observatory, 933 N Cherry Ave., Tucson, AZ 85721 USA}
\author{Brian Svoboda}
\affiliation{National Radio Astronomy Observatory, PO Box O, Socorro, NM 87801, USA}
\affiliation{Jansky Fellow of the National Radio Astronomy Observatory.}
\author{Vianney Taquet}
\affiliation{INAF, Osservatorio Astrofisico di Arcetri, Largo E. Fermi 5, I-50125, Firenze, Italy}
\author{Leonardo Testi}
\affiliation{European Southern Observatory, Karl-Schwarzschild Str. 2, 85748 Garching bei M\"{u}nchen, Germany}
\affiliation{INAF, Osservatorio Astrofisico di Arcetri, Largo E. Fermi 5, I-50125, Firenze, Italy}
\author{Charlotte Vastel}
\affiliation{IRAP, Universit\'{e} de Toulouse, CNRS, CNES, UPS, Toulouse, France}
\author{Serena Viti}
\affiliation{Department of Physics and Astronomy, University College London, Gower Street, London, WC1E 6BT, UK}
\author{Naoki Watanabe}
\affiliation{Institute of Low Temperature Science, Hokkaido University, N19W8, Kita-ku, Sapporo, Hokkaido 060-0819, Japan}
\author{Yoshimasa Watanabe}
\affiliation{Materials Science and Engineering, College of Engineering, Shibaura Institute of Technology, 3-7-5 Toyosu, Koto-ku, Tokyo 135-8548, Japan}
\author{Arezu Witzel}
\affiliation{Univ. Grenoble Alpes, CNRS, IPAG, 38000 Grenoble, France}
\author{Ci Xue}
\affiliation{Department of Chemistry, University of Virginia, McCormick Road, PO Box 400319, Charlottesville, VA 22904, USA}

\author{Bo Zhao}
\affiliation{Center for Astrochemical Studies, Max-Planck-Institut f\"{u}r extraterrestrische Physik (MPE), Gie$\beta$enbachstr. 1, D-85741 Garching, Germany}
\author{Satoshi Yamamoto}
\affiliation{Department of Physics, The University of Tokyo, 7-3-1, Hongo, Bunkyo-ku, Tokyo 113-0033, Japan}
\affiliation{Research Center for the Early Universe, The University of Tokyo, 7-3-1, Hongo, Bunkyo-ku, Tokyo 113-0033, Japan}




\begin{abstract}

We report a study of the low-mass Class-0 multiple system VLA 1623AB in the Ophiuchus star-forming region, using H$^{13}$CO$^+$ ($J=3-2$), CS ($J=5-4$), and CCH ($N=3-2$) lines as part of the ALMA Large Program FAUST. The analysis of the velocity fields revealed the rotation motion in the envelope and the velocity gradients in the outflows (about 2000 au down to 50 au). We further investigated the rotation of the circum-binary VLA 1623A disk as well as the VLA 1623B disk. 
We found that the minor axis of the circum-binary disk of VLA 1623A is misaligned by about 12 degrees with respect to the large-scale outflow and the rotation axis of the envelope.
In contrast, the minor axis of the circum-binary disk is parallel to the large-scale magnetic field according to previous dust polarization observations, suggesting that the misalignment may be caused by the different directions of the envelope rotation and the magnetic field.  
If the velocity gradient of the outflow is caused by rotation,  the outflow has a constant angular momentum and the launching radius is estimated to be $5-16$ au, although it cannot be ruled out that the velocity gradient is driven by entrainments of the two high-velocity outflows.
Furthermore, we detected for the first time a velocity gradient associated with rotation toward the VLA 16293B disk. The velocity gradient is opposite to the one from the large-scale envelope, outflow, and circum-binary disk. The origin of its opposite gradient is also discussed.

\end{abstract}


\section{Introduction} \label{sec:intro}

Angular momentum plays important roles in star-formation processes \citep[e.g.,][]{bod95}.
Star formation proceeds via gravitational collapse of a dense core with a typical size of $\lesssim0.1$ pc \citep{shu77}.
Because of the angular momentum of the dense core, the accretion material cannot directly fall onto the central protostar, which results in the formation of an accretion disk. The accretion disk evolves into a protoplanetary disk as the envelope dissipates. Planets are expected to form in such a protoplanetary disk.

During gravitational collapse, collimated supersonic ($\sim$ 100 km s$^{-1}$) jets
and wider, slower ($\sim$ 10 km s$^{-1}$) outflows remove the angular momentum from the disk, allowing the accretion to the protostar \citep[e.g.,][]{sne80,fra14}.
Recent ALMA high-resolution observations have succeeded in measuring the outflow rotation and have estimated the launching radii of the outflows \citep[e.g.,][]{bje16,hir17,alv17,lee18,zha18}. 
These observations suggest that the low-velocity outflows launch from a radius of about 2 to 25 au in the disk, although in some cases they could be launched from larger distances beyond the disk edge and coinciding with the location of the centrifugal barrier \citep[see e.g. BHB07-11;][]{alv17}.
A relatively large launching radius for outflows seems to be consistent with a disk wind model \citep{kon00} rather than an X wind model \citep{shu00}.
However, it is still unclear how these outflows are related to the angular momentum of a parent core and disk formation.

The magnetic field is also crucial to launching the outflow, in addition to the rotation motion.
The standard picture of protostar formation shows that the B-field couples with the gas component during the collapse of the protostellar core and becomes parallel to the outflow direction.
In contrast, dust polarization observations have shown that the magnetic field (B-field) in protostellar cores does not correlate with the outflow but are randomly oriented \citep{hul14,hul19}.
The angular momentum is also not aligned with the B field \citep{gal20}.

\begin{figure*}[htbp]
\begin{center}
\includegraphics[width=16.cm,bb=0 0 2653 1243]{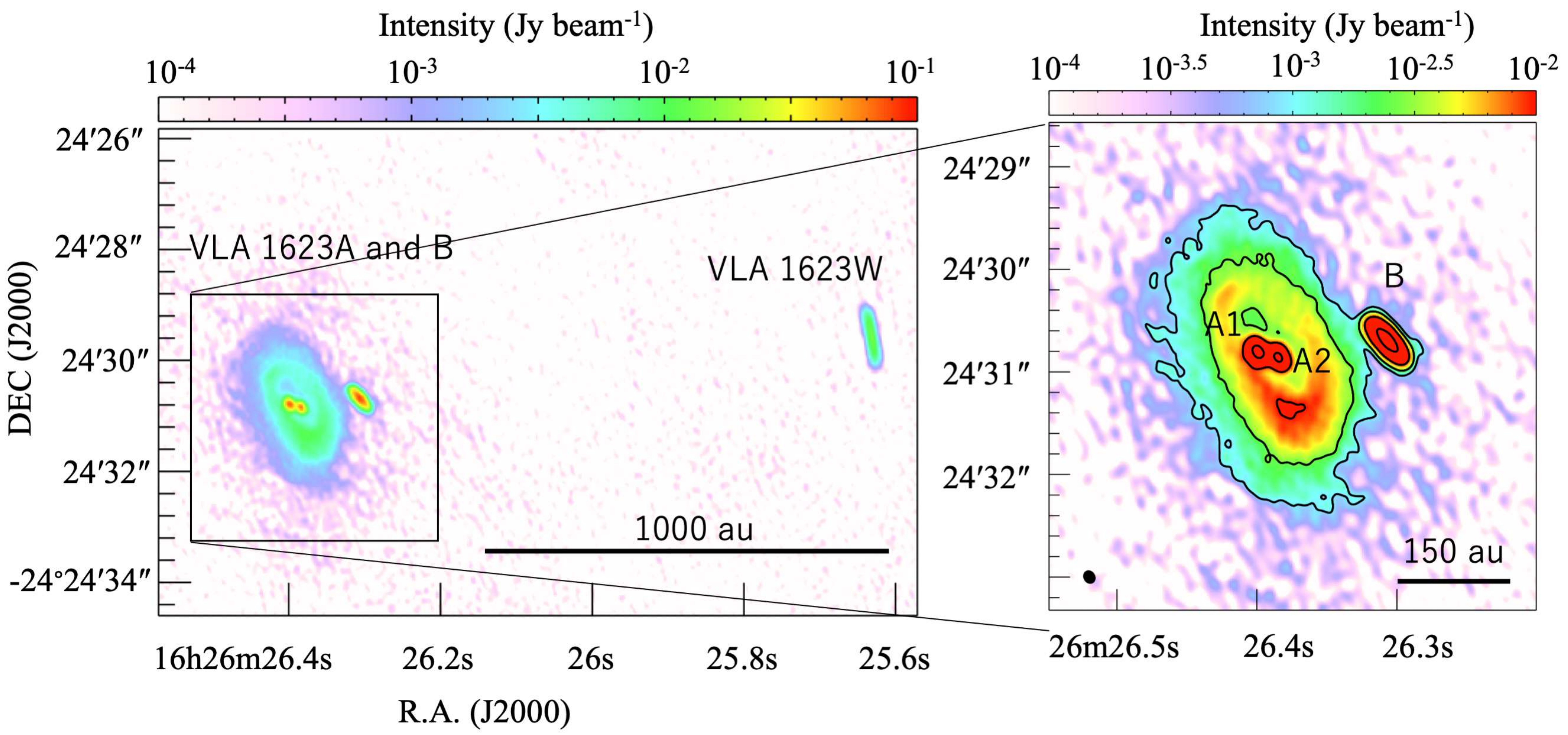}
\end{center}
\caption{Images of the 0.87 mm dust continuum emission observed toward the VLA 1623 region with ALMA. The data is taken from the ALMA archive and details of the observation are described by \cite{har18}. The class 0 source, VLA 1623A, is resolved into two sources, VLA 1623A1 and VLA 1623A2. The beam size is shown in the left bottom corner of the right panel. The contours are 0.93, 2.7, 10, and 50 mJy beam$^{-1}$  ($1\sigma=0.12$ mJy beam$^{-1}$) to indicate the circum-binary disk around VLA 1623A and circum-stellar disk around VLA 1623B.
}
\label{cont}
\end{figure*}

By taking into account the misalignment of the B-field and the angular momentum, ideal and non-ideal magnetohydrodynamic (MHD) simulations have been performed \citep[e.g.,][]{mat04,mac06,hen09,joo12,zhao18,hir19}.\\
\citet{hir19} showed that the misalignment of the B-field and angular momentum causes different directions of an outflow and jet driven from different scales of a circum-stellar disk.
Furthermore, the circum-stellar disk can be warped, as revealed by  the recent discoveries of the warped protostellar disks of HH 212 \citep{cod14}, L1527 \citep{sak19}, and L1489 \citep{sai20}.

To understand the star formation processes, in particular the role of angular momentum, we investigate the rotation motion on various scales from the envelope to the disks in the VLA 1623$-$2417 star-forming system.
VLA 1623$-$2417 (hereafter VLA 1623) is a deeply-embedded protostellar system in Ophiuchus A  \citep{and90} at a distance of $131\pm1$ pc  \citep{gag18}.

VLA 1623 is considered a prototypical Class 0 source \citep{and93}. 
A prominent and complex outflow is observed in CO and H$_{\rm 2}$ emission \citep{and90,den95,yu97,car06}.
Observations at centimeter wavelengths resolved four compact sources, VLA 1623A, VLA 1623B (Knot A), VLA 1623W (Knot B) and Knot C \citep{bon97}.
The nature of VLA 1623A is proposed as the driving source of the outflow and an asymmetric jet composed of VLA 1623B and W \citep{bon97,mau12,har20}, or as protostars \citep{loo00,mur13b}.
The presence of multiple outflows was highlighted in early work \citep{den95,car06}, suggesting that the second scenario was more likely.
Based on the spectral energy distribution (SED) for each component, including 0.42 mm (ALMA Band 9), \textit{Herschel Space Observatory} PACS fluxes and Spitzer IRS spectra, confirms all three sources to be protostellar, with VLA 1623 A and B being Class 0, while VLA 1623W is Class I \citep{mur18}.

Recent high-angular resolution dust continuum observations with ALMA revealed the complicated structure of VLA 1623, including the binarity of VLA 1623A {\citep{har18}.
Figure \ref{cont} shows the 0.87 mm dust continuum image obtained from the ALMA archive.
Details of the observations are described in \citet{har18}.
The VLA 1623A binary, with components A1 and A2 separated by $\sim$ 28 au, is associated with a circum-binary disk. 
It is unclear whether the prominent outflow from VLA 1623A is driven by one or both of the sources in the binary.
VLA 1623B is a very bright source showing a disk-like structure, with a projected separation from VLA 1623A of $\sim$ 150 au to the west. 
A compact but strongly collimated outflow driven by VLA 1623B was detected with ALMA Cycle 0 observations (Santangelo et al. 2015).
VLA 1623B is characterized by presenting a variable H$_{\rm 2}$O maser \citep{fur03}, and SO emission \citep{mur13,hsi20}, but lacking emission in CO isotopologues or DCO$^{\rm +}$ \citep{mur13,mur15}.
Modelling of the C$^{\rm 18}$O emission around VLA 1623A found a rotationally supported disk with $r_{\rm disk}$ $\sim$ 200 au, showing Keplerian rotation out to at least 160 au and an unresolved central mass of $\sim$ 0.2 $M_{\odot}$ \citep{mur13}.
Subsequently, \citet{hsi20} derived a total binary mass of $0.3-0.5$ $M_{\odot}$ using better spatial resolution ALMA data.
Given the large size of the VLA 1623A disk and no apparent distortions at the edge (e.g., disk truncation or heating), the separation between VLA 1623A and B is likely larger than 200 au.
Due to these interesting features of the early stage and its multiplicity, VLA 1623 was selected as a target of the ALMA Large Program FAUST (Fifty AU STudy of the chemistry in the disk/envelope system of Solar-like protostars \footnote{\url{http://faust-alma.riken.jp}}). This program aims at revealing the physical and chemical structure of thirteen nearby protostars ($d=137-235$ pc), at scales from a few 1000 au down to 50 au, by observing various molecular lines.

In this paper, we present a subset of the VLA 1623 results from the FAUST project based on the analysis of the velocity maps of the H$^{13}$CO$^+$ ($J=3-2$), CS ($J=5-4$), and CCH ($N=3-2$) emission.
These molecular lines trace the envelope, outflow, and disk structures of VLA 1623.
In Section \ref{sec:obs} we summarize our observations. Section \ref{sec:res} shows the results and describes the structures of the envelope, outflow, circum-binary disk, and circum-stellar disk. We investigate the rotation motion of the envelope and the outflow, and show the misalignment of their rotation axes on different scales in Section \ref{sec:dis}.
Then, we discuss the origin of the misalignments in Section \ref{sec:mis}. Finally, we summarize the paper in Section \ref{sec:sum}.

\section{Observations} \label{sec:obs}

Single-field observations for VLA 1623 were carried out between November 2018 and April 2019 as part of the ALMA Large Program FAUST.
The molecular lines in Band 6 analyzed in this study are listed in Table \ref{table1}. 
We used the 12-m array data from two different configurations (C43-5 and C43-2 for sparse and compact configurations, respectively) and the 7-m array data of the Atacama Compact Array (ACA/Morita Array), combining these visibility data in the UV plane during the imaging.
In total, the baseline lengths range from 7.2 m to 737.4 m.
The adopted field center was ($\alpha_{2000}$, $\delta_{2000}$)= (16$^{\rm h}$26$^{\rm m}$26\fs392, $-$24\arcdeg 24\arcmin 30\farcs718), which is close to the protostellar position.
The backend correlator for the molecular line observations was set to a resolution of 122 kHz ($\sim0.17$ km s$^{-1}$) and a bandwidth of 62.5 MHz ($\sim 87$ km s$^{-1}$).
The data were reduced in Common Astronomy Software Applications package (CASA) 5.6.1-8 \citep{mcm07} using a modified version of the ALMA calibration pipeline and an additional in-house calibration routine  to correct for the $T_{\rm sys}$ and spectral line data normalization\footnote{\url{https://help.almascience.org/kb/articles/what-errors-could-originate-from-the-correlator-spectral-normalization-and-tsys-calibration}}. 
Self-calibration was carried out using line-free continuum emission for each configuration. The complex gain corrections derived from the self-calibration were then applied to all channels in the data, and the continuum model derived from the self-calibration was subtracted from the data to produce continuum-subtracted line data. A self-calibration technique was also used to align both amplitudes and phases across the multiple configurations.
Images were prepared by using the {\it tclean} task in CASA 5.6.1, where Briggs weighting with a robustness parameter of 0.5 was employed. 
The primary beam correction was applied to all the images presented in this paper.
The maximum recoverable scale ($\theta_{\rm MRS}$) is estimated as $\sim0.6\lambda/L_{\rm  min}$, where $\lambda$ is the observing wavelength and $L_{\rm min}$ is the minimum baseline.
Since the maximum recoverable scales are $\sim20''$, any structures extended more than that size will be resolved out.
The root mean square (r.m.s.) noise levels for H$^{13}$CO$^+$, CS, and CCH are 2.5, 2.3, and  2.8 mJy beam$^{-1}$ channel$^{-1}$, respectively. 
The achieving synthesized beam are shown in Table \ref{line}.

 To understand the relationship between the molecular gas distribution and the protostars we compare the 0.87 mm continuum image previously reported by \citet{har18} with the molecular line distributions.  
The absolute position in the 0.87 mm observation is fixed by comparing the VLA 1623B position of the present FAUST observations at 1.3 mm with the previous 0.87 mm continuum observations. We find that the peak position of VLA 1623B in the 0.87 mm continuum image is offset by $\sim0\farcs1$ from that in the 1.3 mm continuum image. 
This is plausibly due to the proper motion of the source, which will be discussed in a forthcoming study. Even though the offset is smaller than our spatial resolution of $\sim0\farcs5$, it is corrected by shifting the 0.87 mm continuum image in order to compare it with the molecular distributions obtained in this study. 
The offset in VLA 1623A is hardly identified because our spatial resolution is not enough to resolve the binary protostars of A1 and A2.

\begin{table*}[ht]
\caption{ List of Observed Lines $^a$ \label{line}}
\scalebox{0.9}{
\begin{tabular}{lccccccc}
\hline \hline
 Molecule$^a$&Transition & Frequency & $S \mu^2$ & $E_{\rm u}$$k^{-1} $  & Beam size  & r.m.s \\
             &           &  (GHz)    &   ($D^2$) & $(\rm K)$             &             &  (mJy beam$^{-1}$ channel$^{-1}$) \\
 \hline
H$^{13}$CO$^+$ &$ J=3-2$	&      260.25534    &45.6    & 25.0  &0\farcs498$\times$0\farcs469 (P.A. 80$^{\circ}$)  & 2.5  \\   
CS &$J=5-4$&   244.93556    &19.2   &  35.3 &               0\farcs529$\times$0\farcs495 (P.A. 82$^{\circ}$)  &  2.3 \\
CCH & $N=3-2,J=7/2-5/2, F=4-3$ & 262.00426 &   2.3   &  25.2   &0\farcs493$\times$0\farcs465 (P.A. 81$^{\circ}$)  & 2.8 \\	            
CCH & $N=3-2,J=7/2-5/2, F=3-2$& 262.00648  & 1.7   &  25.1      &0\farcs493$\times$0\farcs465 (P.A. 81$^{\circ}$)  & 2.8  \\
\hline
\label{table1}
\end{tabular}}
\begin{flushleft}
\tablecomments{
$^a$ Line parameters are taken from CDMS \citep{edr16}.
}
\end{flushleft}
\end{table*}

\section{Results} \label{sec:res}

In this section, we show the intensity distributions of the molecular line emission.
Figure \ref{3color} shows a three color image of the integrated intensities of H$^{13}$CO$^+$ (blue), CCH (green), and CS (red) emission overlaid on the 0.87 mm dust continuum emission.
The velocity range for integration is set to be $V_{\rm LSR} = -6$ to $10$ km s$^{-1}$.
The detailed distributions of each molecular line are described in below, in Section \ref{subsec:int_maps}.
Based on Figure \ref{3color}, the molecular lines are classified into two groups; the dense gas tracer H$^{13}$CO$^+$ \citep[e.g., ][]{cas02,oni02,ike07} and the outflow cavity tracers CS and CCH \citep[e.g.,][]{zha18} in this source. The CS and CCH molecules are expected to be abundant in (UV-irradiated) warm region even though the critical densities ($n_{\rm cr}$) of these molecular lines are almost similar ($n_{\rm cr}\sim10^{7}$ cm$^{-3}$).

\begin{figure*}[htbp]
\begin{center}
\includegraphics[width=16.cm,bb=0 0 1654 995]{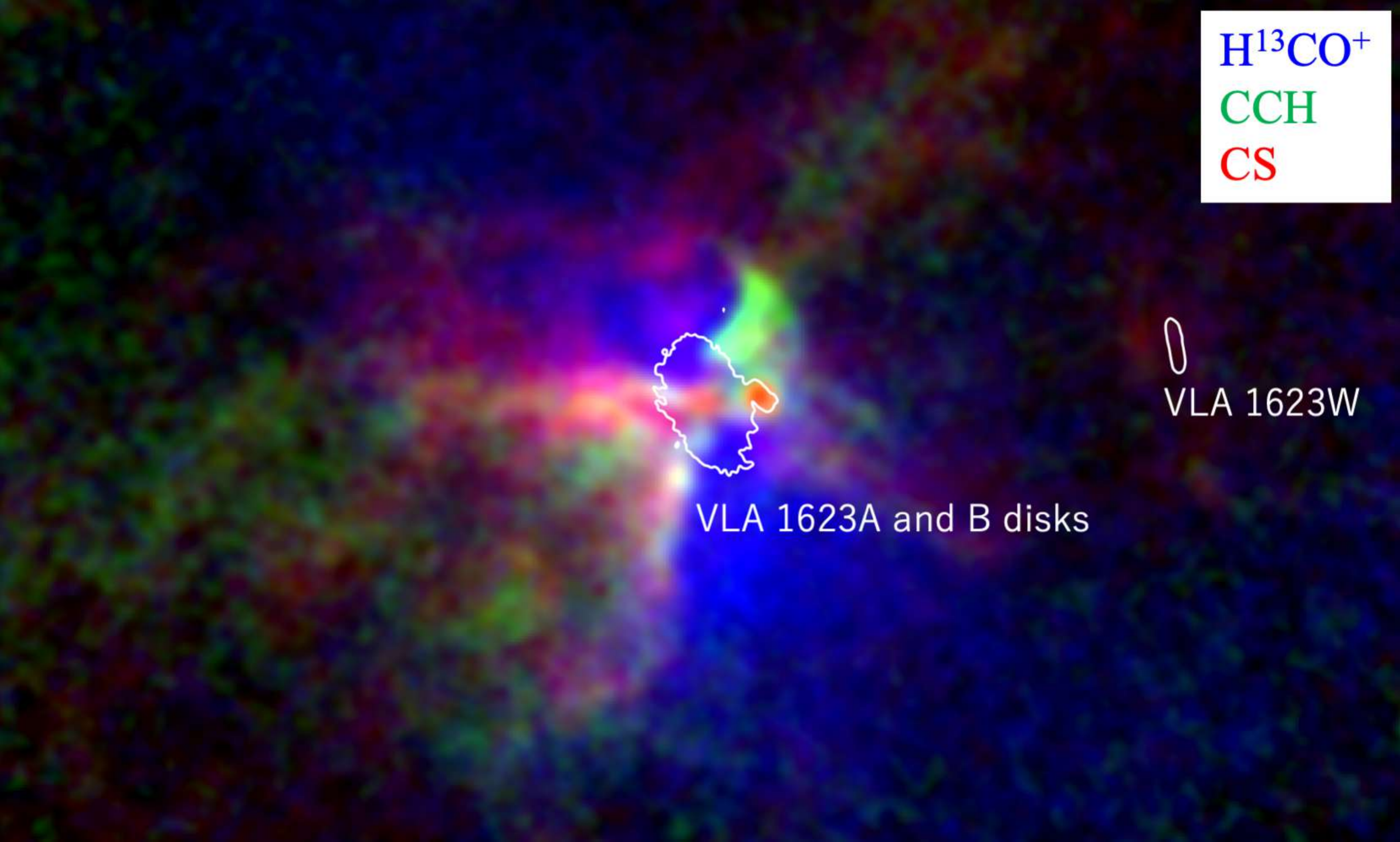}
\end{center}
\caption{A three color image of the integrated intensities of H$^{13}$CO$^+$ (blue), CCH (green), and CS (red) emission toward the VLA 1623 region. The velocity range for integration is  $V_{\rm LSR} = -6$ to $10$ km s$^{-1}$. The white contour shows the 0.87 mm dust continuum emission presented by \citet{har18} and shown in Figure \ref{cont} to indicate the disk and protostellar positions.
}
\label{3color}
\end{figure*}

\subsection{Spectra of the Molecular Lines toward the Protostars}\label{subsec:spec}

Figures \ref{spectra1} and \ref{spectra2} show the spectral line profiles toward the protostellar positions of VLA 1623A1 and B. Note that our observations cannot resolve the positions of VLA 1623A1 and A2.  The CCH molecule has hyperfine structure, with two transitions separated by 2.2 MHz, corresponding to $\sim2.5$ km s$^{-1}$. The LSR velocity scale is defined relative to the lower frequency ($N=3-2,J=7/2-5/2, F=4-3$) CCH transition.

The most prominent feature of these spectra is recognized as absorptions at the systemic cloud velocity of $V_{\rm LSR}=3.8$ km s$^{-1}$ \citep{nar06}.
In particular, the absorption feature of the H$^{13}$CO$^+$ and CCH lines is found toward each protostar position, suggesting that the lines are absorbed against the optically-thick continuum background of the protostellar disk(s) and/or self-absorption caused by the foreground (infalling) gas.
 These absorption features are also observed in the channel maps of the H$^{13}$CO$^+$, CS, and CCH emission (see the following subsections).
We discuss the origin of the absorption feature in Section \ref{subsec:abs}.

When we assume that the absorption is originated from the optically thick background emission, the systemic velocity in front of these protostars is determined to be 3.8 km s$^{-1}$, which is consistent with that of the parent protostellar core \citep{nar06}.
Even if the absorption is due to the self-absorption or infall motion, the systemic velocity would not be much different from 3.8 km s$^{-1}$.

Another interesting feature is found in the CS line toward VLA 1623B (Figure \ref{spectra1} bottom panel).
The CS emission is detected in a wider velocity range of $-6$ to $12$ km s$^{-1}$, while the H$^{13}$CO$^+$ and CCH emission is detected in a narrower velocity range of  $2.5$ to $5$ km s$^{-1}$ (per hyperfine component).
We discuss these CS high velocity components in terms of a rotation motion of the VLA 1623B disk in Section \ref{subsec:rot_diskb}.

\begin{figure}[htbp]
\includegraphics[width=8.5cm,bb=0 0 1804 1135]{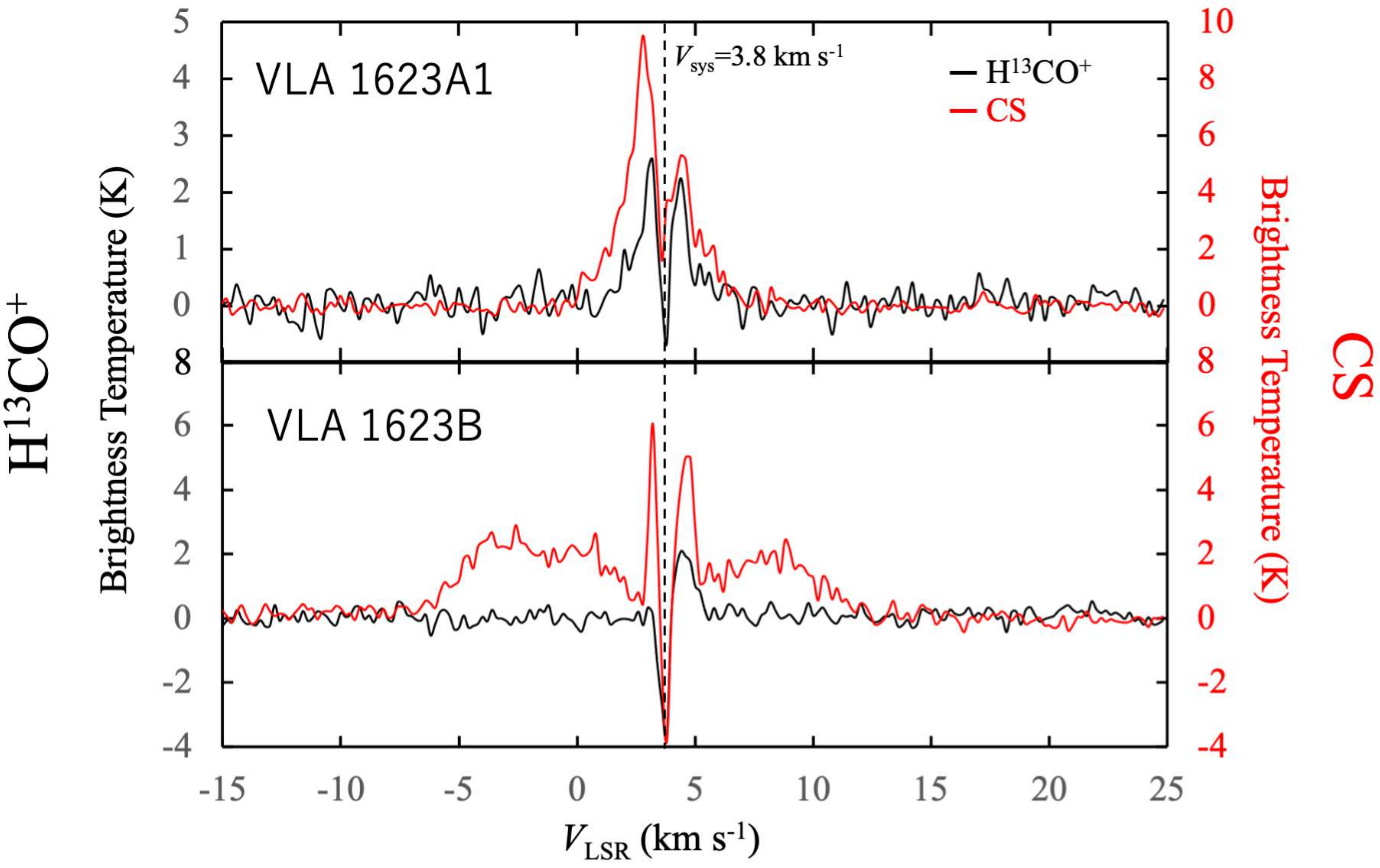}
\caption{ H$^{13}$CO$^+$ ($J=3-2$) and CS ($J=5-4$) spectra toward the protostellar positions of the VLA 1623A1 and B are shown as the black and red lines, respectively. Vertical dashed lines stand for the source systemic velocity ($3.8$ km s$^{-1}$).
}
\label{spectra1}
\end{figure}

\begin{figure}[htbp]
\includegraphics[width=8.cm,bb=0 0 1456 1138]{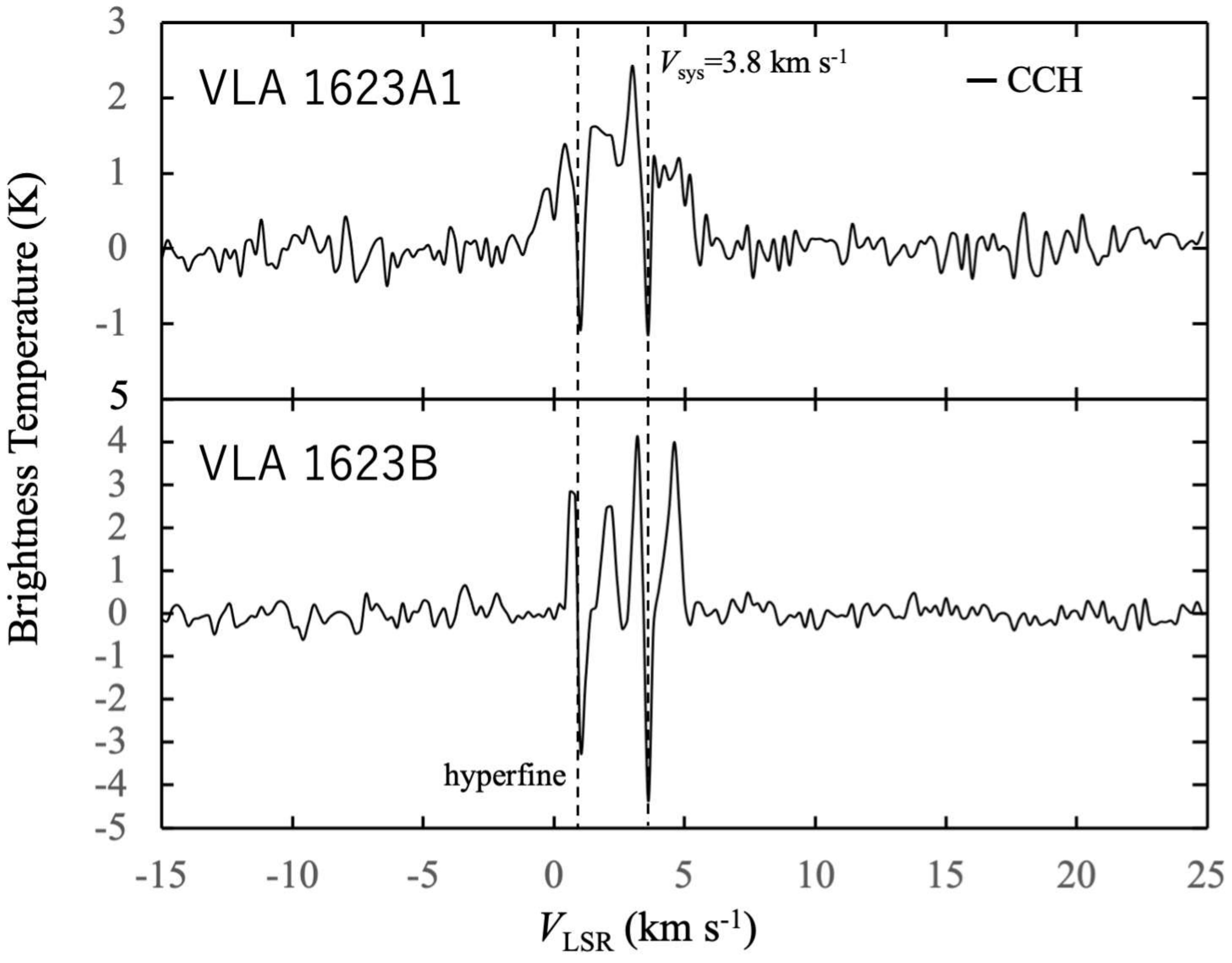}
\caption{ CCH ($N=3-2,J=7/2-5/2, F=4-3$) and ($N=3-2,J=7/2-5/2, F=3-2$) spectra toward the protostellar positions of the VLA 1623A1 and B are shown. Vertical dashed lines stand for the source systemic velocity ($3.8$ km s$^{-1}$) relative to the $F=4-3$ and $F=3-2$ hyperfine components. The $V_{\rm LSR}$ is calculated by using the $F=4-3$ hyperfine component.
}
\label{spectra2}
\end{figure}

\subsection{The velocity-integrated intensity maps}\label{subsec:int_maps}

In this section, we describe the distributions of the H$^{13}$CO$^+$, CS, and CCH emission from the velocity-integrated intensity maps.

\subsubsection{The Envelope Structure shown by the H$^{13}$CO$^+$ emission}\label{subsec:h13cop}

\begin{figure*}[htbp]
\includegraphics[width=16.cm,bb=0 0 2660 1404]{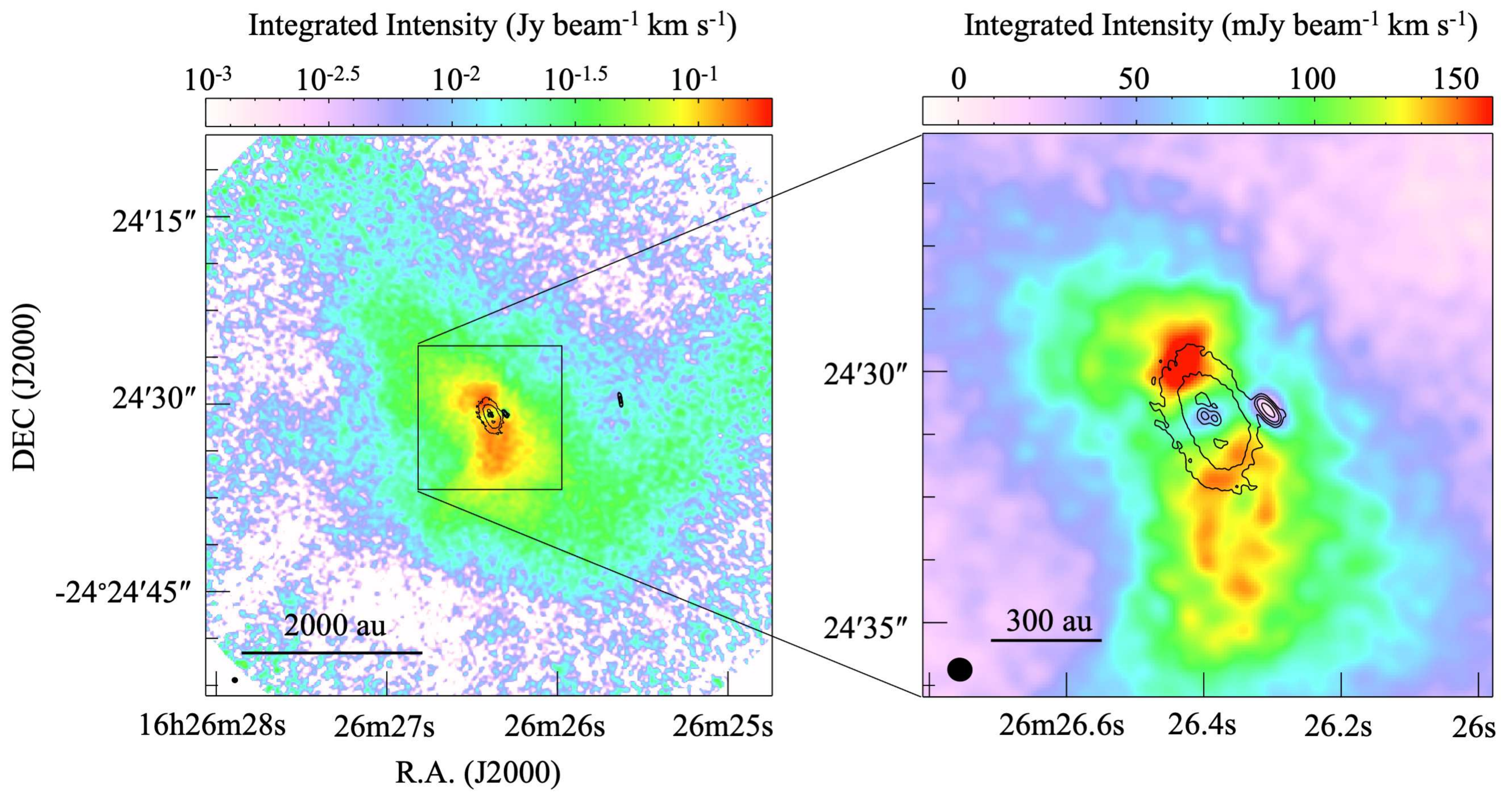}
\caption{Integrated intensity maps of the H$^{13}$CO$^+$ emission toward the VLA 1623 region. The velocity range for integration is $1.4-6.0$ km s$^{-1}$.
The 1-sigma noise level is $2.9\times10^{-3}$ Jy beam$^{-1}$ km s$^{-1}$.
The black  contours show the 0.87 mm dust continuum  emission presented by \citet{har18} and shown in Figure \ref{cont}, to indicate the disk and protostellar positions. The beam size of the H$^{13}$CO$^+$ emission is shown in the bottom left corner of each panel. 
}
\label{h13cop_mom0}
\end{figure*}

Figure \ref{h13cop_mom0} shows the integrated intensity map of the H$^{13}$CO$^+$ emission overlaid on the 0.87 mm dust continuum emission tracing the circum-binary disk of VLA1623A and the protostellar disk around VLA1623B.
The H$^{13}$CO$^+$ emission is widely detected and shows an elongated structure along the northeast to southwest direction.

The size of the H$^{13}$CO$^+$ distribution is about 4,000 au, corresponding to 0.02 pc, which is similar to the typical size of compact dense cores in various star-forming regions including high-mass star forming regions in Infrared Dark Clouds (IRDCs) \citep[e.g.,][]{oha16}.
Therefore, the distribution of the H$^{13}$CO$^+$ emission appears to represent the dense core in the VLA 1623 region.

Figure \ref{h13cop_mom0} (right panel) also shows the zoomed-in image around the circum-binary disk and protostars.
Lower integrated intensities are found toward the protostars of VLA 1623A1, A2, and B.
In particular, an absorption feature ($\sim -0.05$ Jy beam$^{-1}$, corresponding to $\sim4$ K at $V_{\rm LSR}=3.8$ km s$^{-1}$) is found toward VLA 1623B as shown in Figure \ref{spectra1}.

\subsubsection{The Outflow Cavity Structure shown by the CS and CCH emission}\label{subsec:cs_cch}

\begin{figure*}[htbp]
\includegraphics[width=16.cm,bb=0 0 2648 1398]{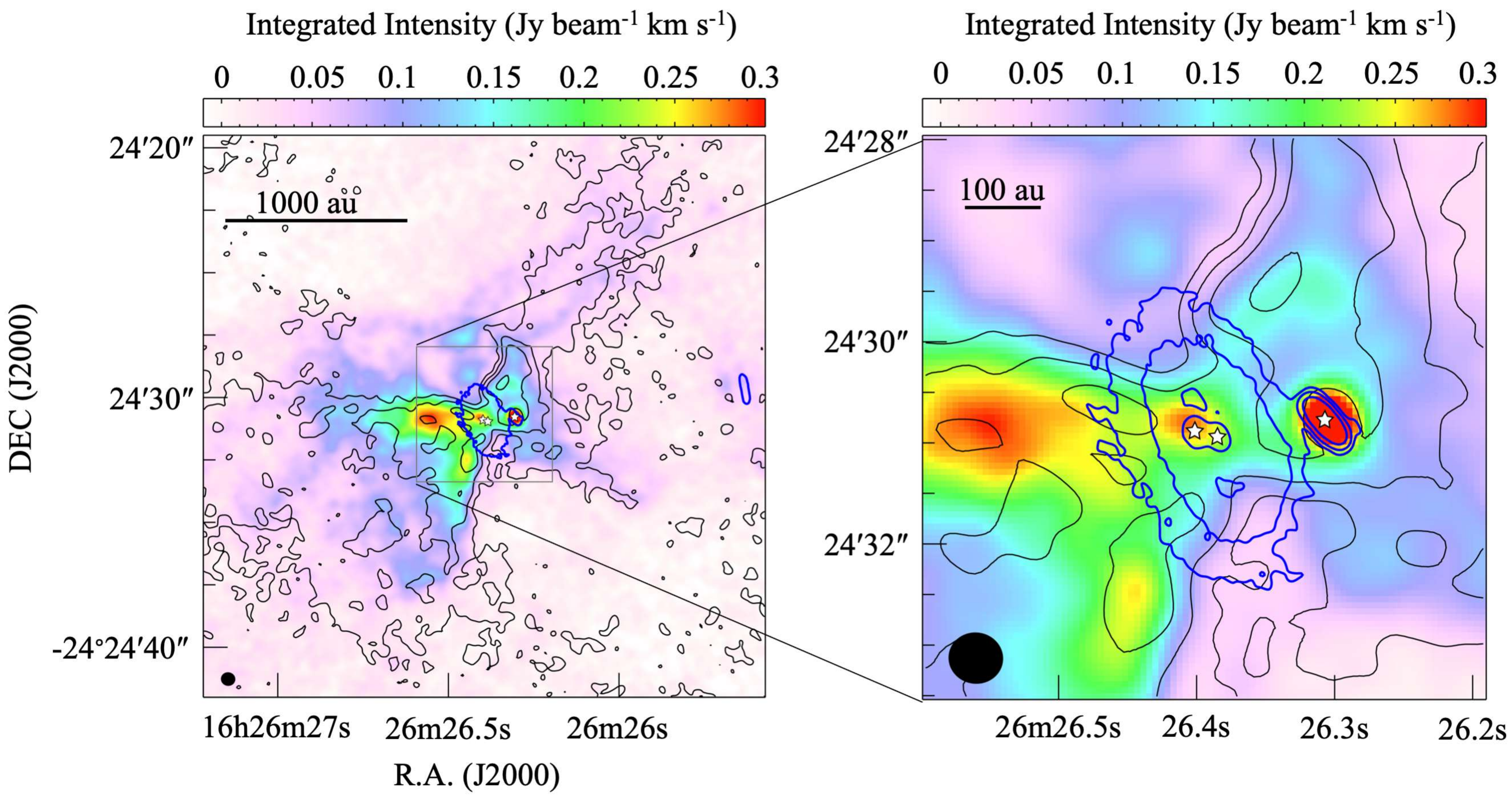}
\caption{Integrated intensity maps of the CS emission in color scale and the CCH emission in black contours toward the VLA 1623 region. The velocity range for integration is $-6$ to $+10$ km s$^{-1}$.
The contours of the CCH emission are $0.02, 0.05, 0.1,$ and  $0.2$ Jy beam $^{-1}$ km s$^{-1}$, where $1\sigma=0.0025$ Jy beam$^{-1}$ km s$^{-1}$.
The blue thick contours show the 0.87 mm dust continuum  emission \citep{har18} to indicate the circum-binary disk and protostellar positions. The beam size of the CS emission  is shown in the bottom-left corners.
}
\label{cs_cch_mom0}
\end{figure*}

Figure \ref{cs_cch_mom0} shows the integrated intensity maps of the CS and CCH emission. The color image indicates the CS distribution, while the black contours indicate the CCH emission distribution. The blue contours show the 0.87 mm dust continuum emission to indicate the circum-binary disk and protostellar positions.
The zoom up view around the protostars is also shown in the right panel.

The distributions of CS and CCH are similar to each other, and show ``X-shaped'' structures, indicating the edges of the outflow cavities.
The outflow cavity structure is elongated in the southeast-northwest direction, which is consistent with the direction of the CO outflow \citep{san15,har20}.
 We discuss the outflow properties in Section \ref{subsec:outflow}.

The distributions of CS and CCH are locally different in some parts, even though the CS and CCH distributions reveal a similar outflow cavity structures at large scales.
The most prominent difference is seen in the disk around VLA 1623B (see right panel of Figure \ref{cs_cch_mom0}).
The integrated intensity of the CS emission is peaked at this position, while that of the CCH emission shows a hole structure.
A similar feature of the CCH intensity depression near the protostar is shown by \citet{oya17} in L483.
They suggested that the depression of the CCH emission is caused by the gas-phase destruction and/or depletion onto dust grains.
We suggest that the CS emission in VLA 1623B comes from the high velocity components of a rotation motion (Figure \ref{spectra1} and see Section \ref{subsec:rot_diskb} for further discussion).
In contrast, the CCH emission is not detected in the high velocity regions, and is absorbed at the systemic velocity ($V_{\rm LSR}\sim3.8$ km s$^{-1}$). Note that the CS emission also shows the absorption feature around the systemic velocity.

Furthermore, other different distributions are also found in the outflow cavity structures.
For example, the CS outflow cavity structure is slightly wider compared to the CCH outflow cavity in the northern part of the southeast outflow.
In addition, the CS emission is detected in the surrounding regions perpendicular to the outflow direction.

To conclude, based on the reported spatial distributions, we suggest the CS emission traces not only the outflow cavity structure but  also the disk rotation and accretion flow as similar to the binary Class 0 source of IRAS 16293 A1/A2 \citep{mau20}. 
In contrast, the CCH emission mainly shows the outflow cavity structure.

\subsection{Analysis of the Channel Maps}\label{subsec:an_cm}

To investigate the kinematic motions of the envelope, outflow, and disks, we show the channel maps of the H$^{13}$CO$^+$, CS,  and  CCH emission.
The velocity range for the integration and channel step are set to be 0.6 km s$^{-1}$ in each channel map.

\subsubsection{H$^{13}$CO$^+$ Channel Maps}\label{subsubsec:cm_h13cop}

\begin{figure*}[htbp]
\includegraphics[width=16.cm,bb=0 0 2688 1568]{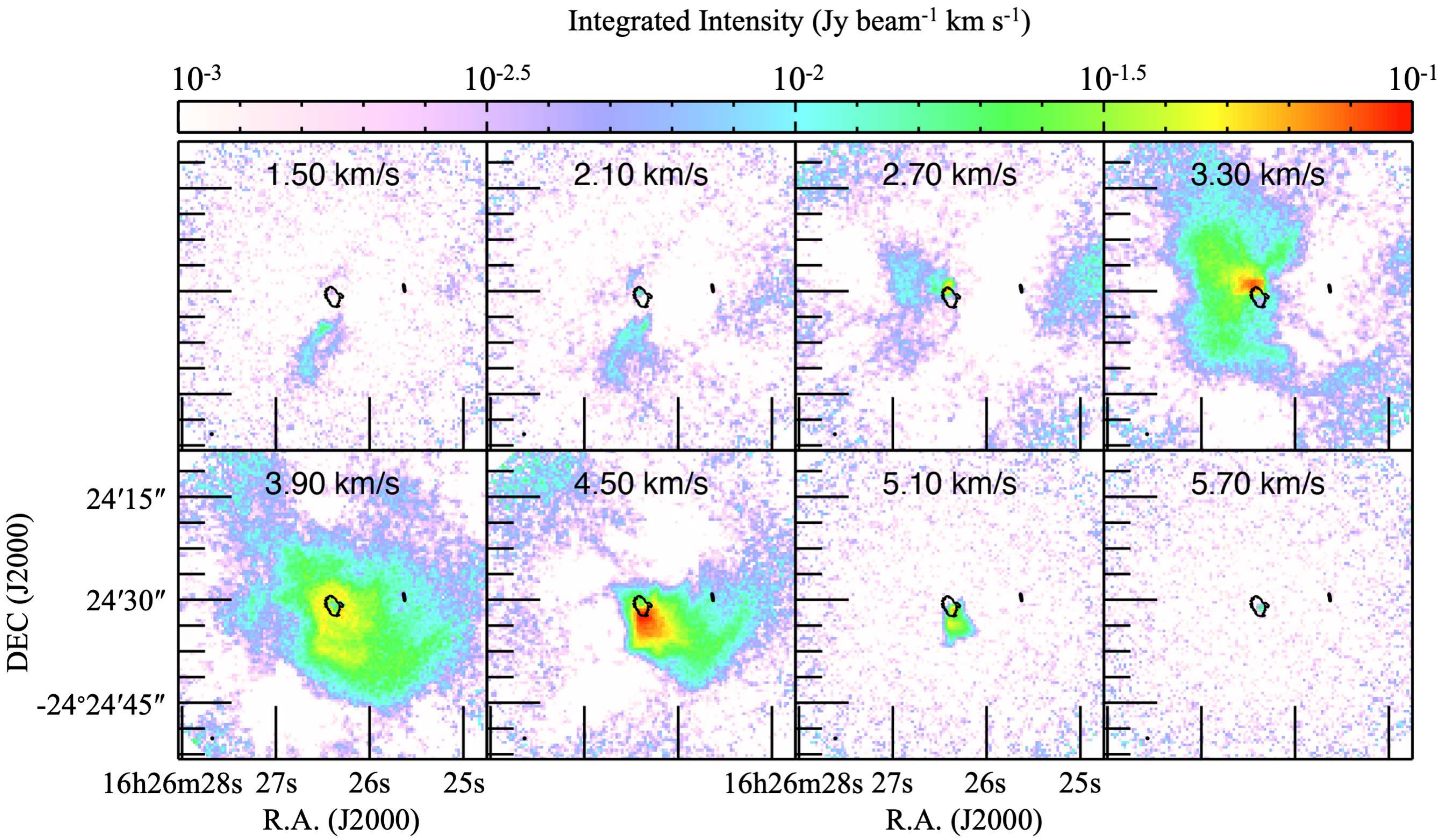}
\caption{Channel maps of the H$^{13}$CO$^+$ ($J=3-2$) emission, overlaid with the 0.87 mm dust continuum emission \citep{har18} as black contours. The contour is 0.93 mJy beam$^{-1}$, indicating the disks of VLA 1623A, B, and W. The channel width and step are 0.6 km s$^{-1}$, and the channel centroid velocities are labeled at the upper center of each panel.  
The 1-sigma noise level for H$^{13}$CO$^+$ is $1.2\times10^{-3}$ Jy beam$^{-1}$ km s$^{-1}$ in each panel. The beam size is shown in the bottom-left corners.
}
\label{h13cop_channel_map}
\end{figure*}

\begin{figure*}[htbp]
\includegraphics[width=16.cm,bb=0 0 2684 1563]{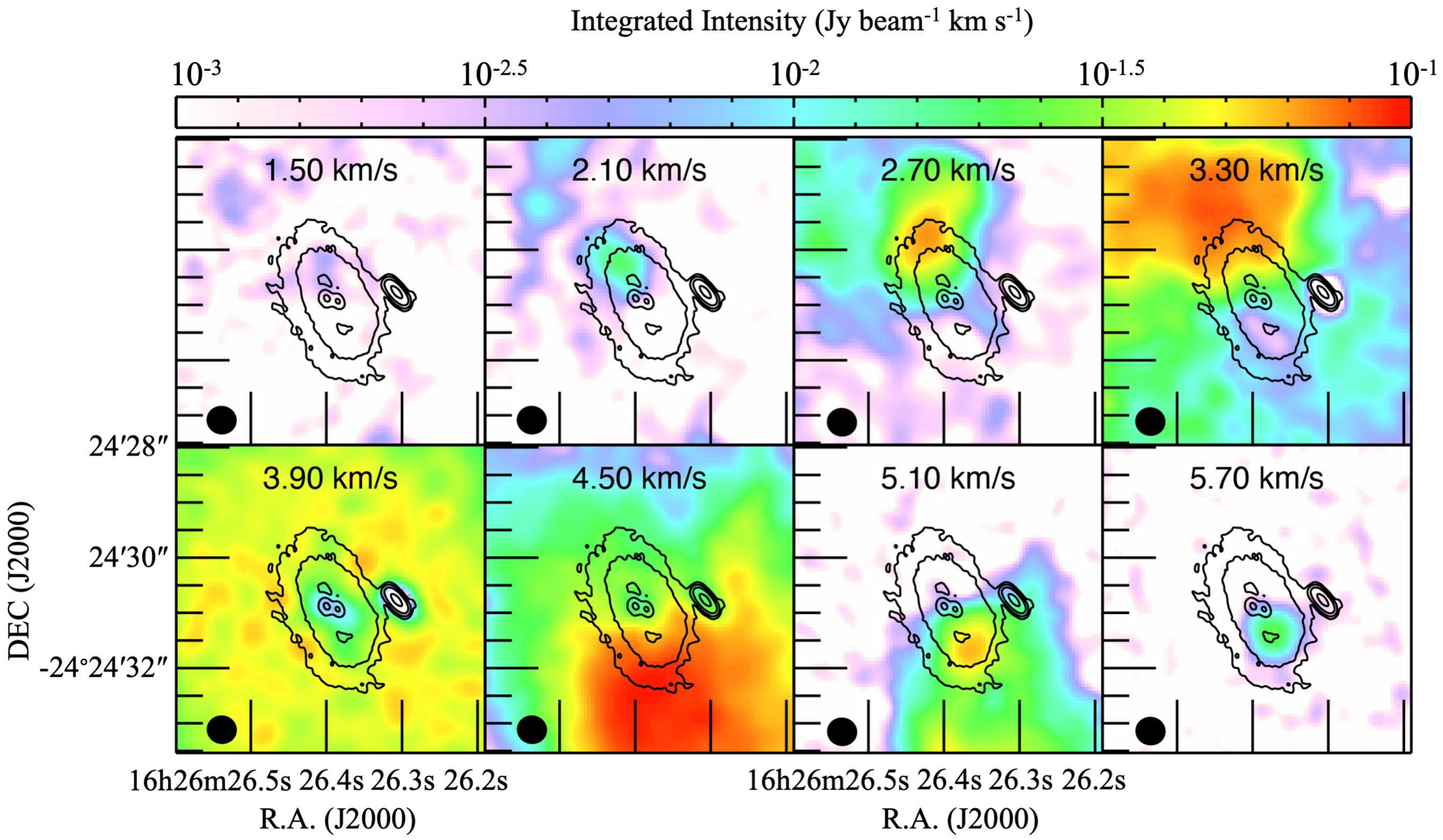}
\caption{The same channel maps of the H$^{13}$CO$^+$ ($J=3-2$) emission as Figure \ref{h13cop_channel_map}, but zoomed-in views around the protostars. The black contours indicate the 0.87 mm dust continuum emission \citep{har18} to indicate the disks and stellar positions. The contour levels are the same as in Figure \ref{cont}. The beam size is shown in the bottom-left corners.
}
\label{h13cop_channel_map_zoom}
\end{figure*}

Figure \ref{h13cop_channel_map} shows channel maps of the H$^{13}$CO$^+$ emission to illustrate several components showing the different structures.

In the velocity range $<2.1$ km s$^{-1}$, the H$^{13}$CO$^+$ emission shows a filamentary structure to the south of the protostars.
A similar filamentary structure is also identified by \citet{hsi20} in SO emission in the same velocity range.   {\citet{hsi20}} interpret the SO emission as tracing an accretion flow toward the VLA 1623A and B disks.

In the channel maps with $V_{\rm LSR}=3.3-4.5$ km s$^{-1}$, the H$^{13}$CO$^+$ emission is widely distributed, tracing the envelope structure.
The blue-shifted components are located in the northeast side, while the red components are located in the southwest side, indicating a rotation motion of the envelope.

Figure \ref{h13cop_channel_map_zoom} shows a zoomed-in view around the protostars.
In the channel maps of $V_{\rm LSR}=2.1-2.7$ and $V_{\rm LSR}=5.1-5.7$ km s$^{-1}$, the H$^{13}$CO$^+$ emission is detected around the circum-binary disk.
In particular, the velocity components ($V_{\rm LSR}=2.1$ and 5.7 km s$^{-1}$) are detected inside the circum-binary disk.
The velocity gradient is found to be consistent with the envelope rotation, indicating the rotation of the circum-binary disk.
Therefore, we suggest that H$^{13}$CO$^+$ would be a good tracer for probing the kinematics of the disk \citet[e.g.,][]{tob11}.

The absorption feature is found toward the VLA 1623B disk at the velocity of $3.3-3.9$ km s$^{-1}$.
In addition, the H$^{13}$CO$^+$ emission is depressed toward the protostars of VLA 1623 A1 and A2 in the circum-binary disk.
This drop in emission would be caused by the optically-thick continuum emission of the background sources of the protostellar disks and/or less abundance of the H$^{13}$CO$^+$ molecule.

\subsubsection{CS Channel Maps}\label{subsubsec:cm_cs}

\begin{figure*}[htbp]
\includegraphics[width=16.cm,bb=0 0 2654 1559]{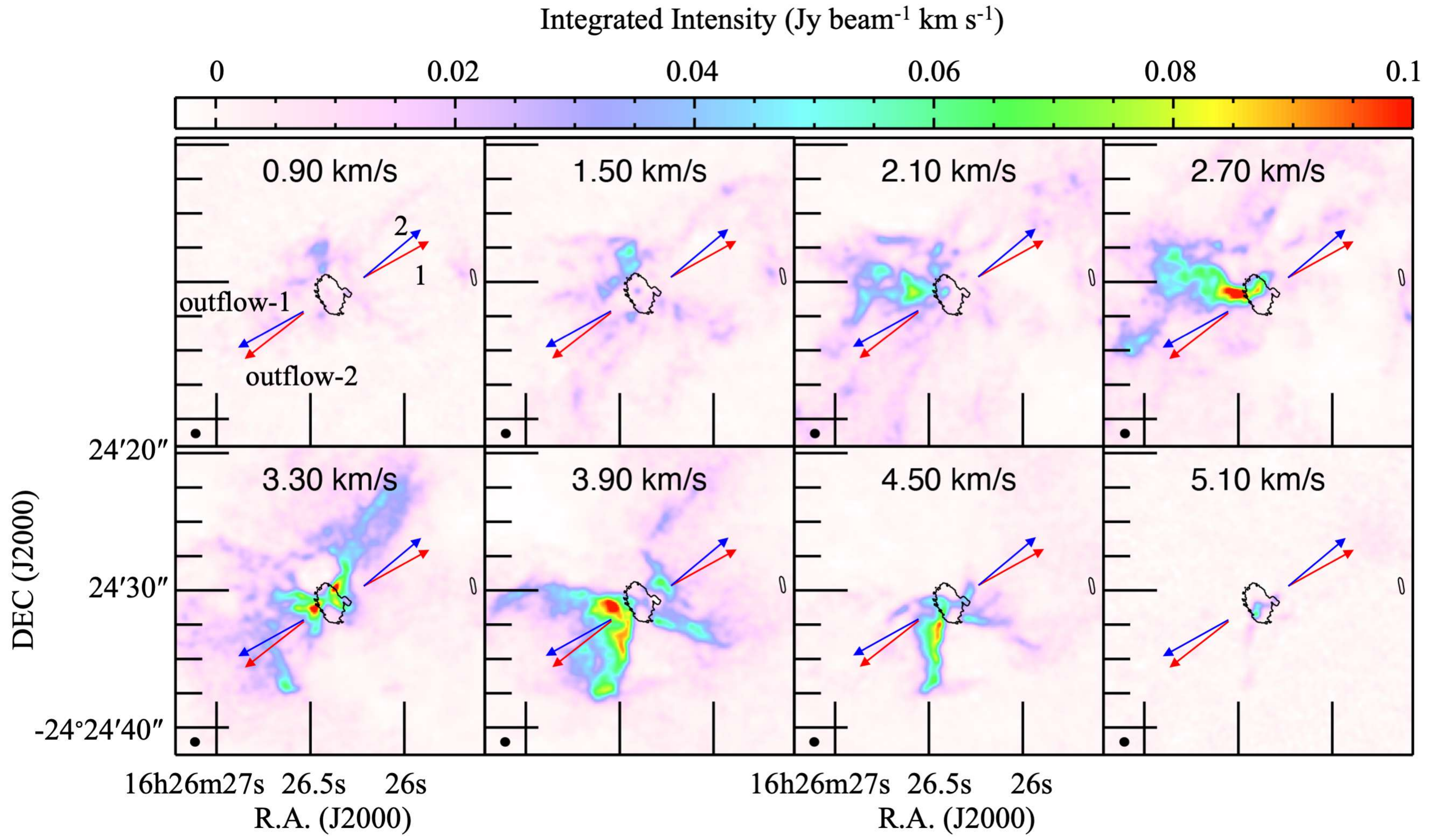}
\caption{Channel maps of the CS emission, overlaid with the 0.87 mm continuum emission \citep{har18} as black contours. The contourr level is 0.93 mJy beam$^{-1}$, indicating the disks of VLA 1623A, B, and W.  The blue and red arrows indicate the directions of the CO outflows observed by \citet{har20}.}
The channel width and step are 0.6 km s$^{-1}$, and the channel centroid velocities are labeled at the upper center of each panel.
The 1-sigma noise level for each panel is  $0.0012$ Jy beam$^{-1}$ km s$^{-1}$. The beam size is shown in the bottom-left corners.

\label{cs_channel_map}
\end{figure*}

Figure \ref{cs_channel_map} shows 0.6 km~s$^{-1}$ channel maps of the CS emission, as for the H$^{13}$CO$^+$ emission.
As shown in the integrated intensity maps of the CS and CCH emission (Figure \ref{cs_cch_mom0}), the CS emission mainly traces the outflow cavity structure.
 We also indicate the directions of the CO outflows observed by \citet{har20} as the blue and red arrows. We find that the outflow cavity encompasses these two high-velocity outflows.

In the channel maps centered at $V_{\rm LSR}=0.9$ and 1.5 km s$^{-1}$, the CS emission shows an extension perpendicular to the outflow.
In particular, the CS emission is found to the north of the circum-binary disk. This would be part of the accretion flow toward the circum-binary disk.

In the channel maps from $V_{\rm LSR}=2.1$ to 5.1 km~s$^{-1}$, the CS emission mainly shows the outflow cavity structure. 
In the channel maps at $V_{\rm LSR}=2.1$ and 2.7 km s$^{-1}$, the northern part of the outflow is detected. In contrast, the southern part of the outflow is detected in the channel maps at $V_{\rm LSR}=4.5$ and 5.1 km s$^{-1}$.

The ``X-shaped'' outflow cavity structure is found in the channel map of $V_{\rm LSR}=3.9$ km s$^{-1}$, which indicates the systemic velocity of the outflow is $\sim3.9$ km s$^{-1}$.
Therefore, the systemic velocities of the envelope and outflow are consistent, to within the spectral resolution of the channel maps.
In addition, the velocity gradient along the outflow was not identified.

No emission is found toward the VLA 1623B disk in the channel maps at $V_{\rm LSR}=2.1$ to 4.5 km~s$^{-1}$ as in the case for the H$^{13}$CO$^+$ emission.
In particular, the absorption is seen in the outflow cavity structure shown in the channel maps at $V_{\rm LSR}=3.3$ to 5.1 km~s$^{-1}$, indicating that the CS emission is absorbed against the optically thick emission of the background continuum source of the VLA 1623B disk. 
In other words, the background dust emission is optically thick and masks the line emission.
This means that the outflow is located in front of the VLA 1623B disk.

\begin{figure*}[htbp]
\includegraphics[width=16.cm,bb=0 0 1958 1545]{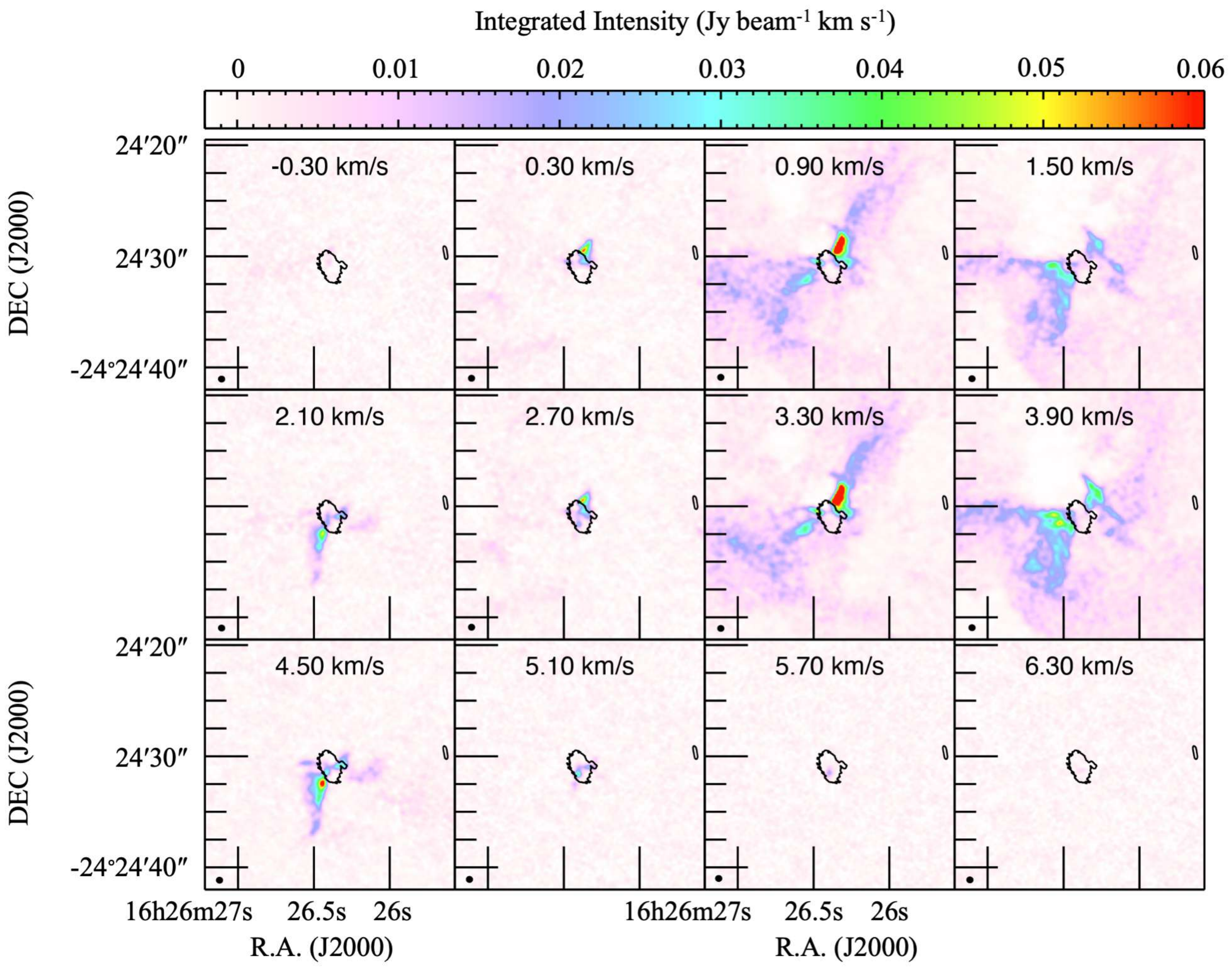}
\caption{Channel maps of the CCH ($N=3-2,J=7/2-5/2, F=4-3$) and ($N=3-2,J=7/2-5/2, F=3-2$) emission, overlaid on the 0.87 mm continuum emission \citep{har18} as black contours. The contour level is 0.93 mJy beam$^{-1}$, indicating the disks of VLA 1623A and B. The channel width and step are 0.6 km s$^{-1}$, and the channel centroid LSR velocities relative to the rest frequency of the $N=3-2, J=7/2-5/2, F=4-3$ transition are labeled at the upper center of each panel.
The 1-sigma noise level for each panel is $0.0014$ Jy beam$^{-1}$ km s$^{-1}$. The beam size is shown in the bottom-left corner of each panel.}
\label{cch_channel_map}
\end{figure*}

\subsubsection{CCH Channel Maps}\label{subsubsec:cm_cch}

Figure \ref{cch_channel_map} shows the channel maps of the CCH emission.
The CCH molecule has hyperfine structure with the two transitions of CCH separated by 2.2 MHz  (see Table \ref{table1}), corresponding to $\sim2.5$ km s$^{-1}$. Therefore, the same distribution is repeated in the channel maps at an interval of 2.5 km s$^{-1}$.

The CCH channel maps show the clear outflow cavity structure.
 The blue shifted emission can be seen in the northern part of the outflow cavity in $V_{\rm LSR}=0.3-0.9$ and  $V_{\rm LSR}=2.7-3.3$, while the red shifted emission is appeared in the southern part of the outflow cavity in  $V_{\rm LSR}=2.1$ and  $V_{\rm LSR}=4.5-5.7$.
This velocity gradient is consistent with the CS channel maps, indicating that the velocity gradient is perpendicular to the outflow lobes.
However, the accretion flow or additional components surrounding the protostars identified in the CS emission is not observed in the CCH emission.
Thus, the CCH emission traces the whole outflow cavity structure, while the CS emission traces not only the outflow cavity but also the disk components and accretion flows.

\section{Discussion} \label{sec:dis}

\subsection{The origin of the absorption feature toward the protostellar positions} \label{subsec:abs}

We found the absorption features at $\sim3.8$ km s$^{-1}$  toward the protostellar positions shown in the spectra (Figures \ref{spectra1}  and \ref{spectra2}) in the channel maps (Figures \ref{h13cop_channel_map}, \ref{cs_channel_map}, and \ref{cch_channel_map}).
If these are caused by self-absorption, the molecular line emission needs to be highly optically thick to produce the absorption feature, irrelevant to the optical depth of the dust continuum emission.
However, we found that the H$^{13}$CO$^+$ and CCH emission is optically thin with $\tau\sim0.1-0.3$ if the continuum background is assumed to be optically thin because the peak brightness temperatures of the emission is $\sim2-10$ K in the protostellar positions.
We assume that the excitation temperature is consistent with the dust temperature of $\sim40$ K, which is derived by the SED fitting toward the VLA 1623A and B positions \mbox{\citep{mur18}}.
Note that the brightness temperatures of the 0.87 mm continuum emission reach as high as $\sim60$ K at those protostellar positions. If we assume the brightness temperature of 60 K as the excitation temperature, the optical depths of the molecular line emission become much lower.
Even if we assume an excitation temperature of 10 K, the H$^{13}$CO$^+$ and CCH emission is still optically thin with $\tau\sim0.3$.
Furthermore, observations with the single-dish telescope of the Caltech Submillimeter Observatory (CSO) showed that the H$^{13}$CO$^+$ ($J=4-3$) emission is optically thin in this region \mbox{\citep{nar06}}.
Therefore, it is likely that the lines are absorbed against the optically thick background emission of the protostellar disks even though the self-absorption and infall motions may also affect the line profiles.
This can be confirmed by the spatial distributions of these lines. The depressions of the H$^{13}$CO$^+$ emission toward VLA 1623B is found across a wide range of velocities, as seen in the channel maps.
Therefore, it is likely that VLA 1623B is highly embedded within its envelope.

Alternatively, the depression toward the protostars of VLA 1623 A1 and A2 could be the result of absorption due to the background emission from the circum-stellar disks around these protostars.
In addition, it might also be possible that the depression is caused by lower abundance of the HCO$^+$ molecule in the dense regions of the circum-stellar disks due to a decrease of the ionization degree.

\subsection{The Rotation Axes of the Envelope and Cicrum-Binary Disk} \label{subsec:rot}

In this subsection, we investigate the rotation motions from the large scales of the envelope and outflow to the small scales of the circum-binary disk and the cirum-stellar disk around VLA 1623AB, and derive their rotation axes.
First, we show the rotation axes of the envelope ($\sim1000$ au) and circum-binary disk  ($\sim100$ au) using the H$^{13}$CO$^+$ emission.
Then, we investigate the outflow direction and rotation by using both CS and CCH emission.

\begin{figure*}[htbp]
\includegraphics[width=16.cm,bb=0 0 2712 1298]{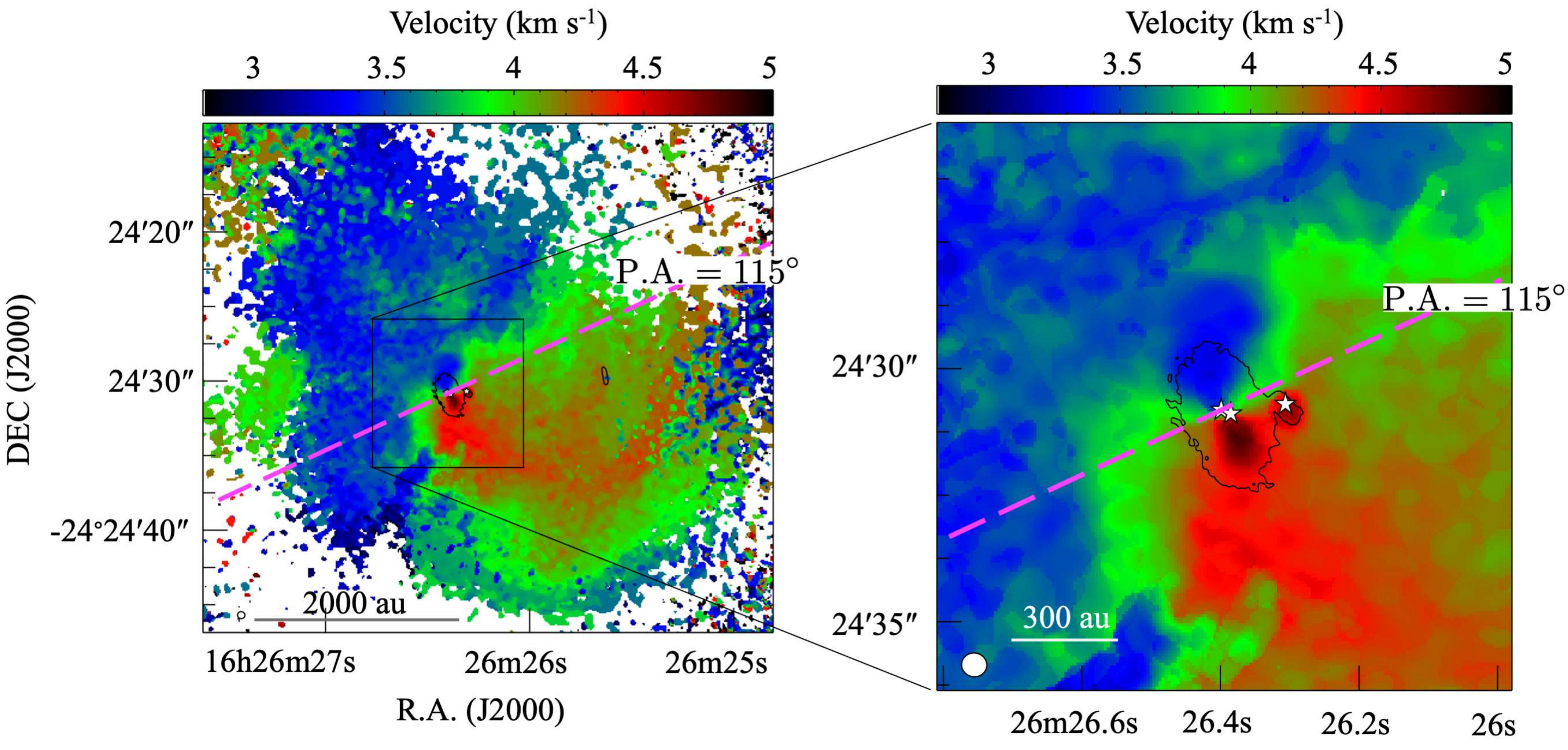}
\caption{Intensity-weighted mean velocity (moment 1) maps of the H$^{13}$CO$^+$ line toward the VLA 1623 region. The right panel is the zoomed-in view around the protostars. The mean velocity is derived using a velocity range of $1.2-6.2$ km s$^{-1}$ with an intensity of $>0.015$ Jy beam$^{-1}$ ($6\sigma$).
The black contour indicates the 0.87 mm dust continuum emission \citep{har18} with an intensity of 0.93 mJy beam$^{-1}$. Magenta dashed lines indicate the minor axis of the  circum-binary disk. The protostellar positions are shown as the star markers. The beam size is shown in the bottom-left corners.
}
\label{h13cop_mom1}
\end{figure*}

\begin{figure*}[htbp]
\includegraphics[width=16.cm,bb=0 0 2712 1298]{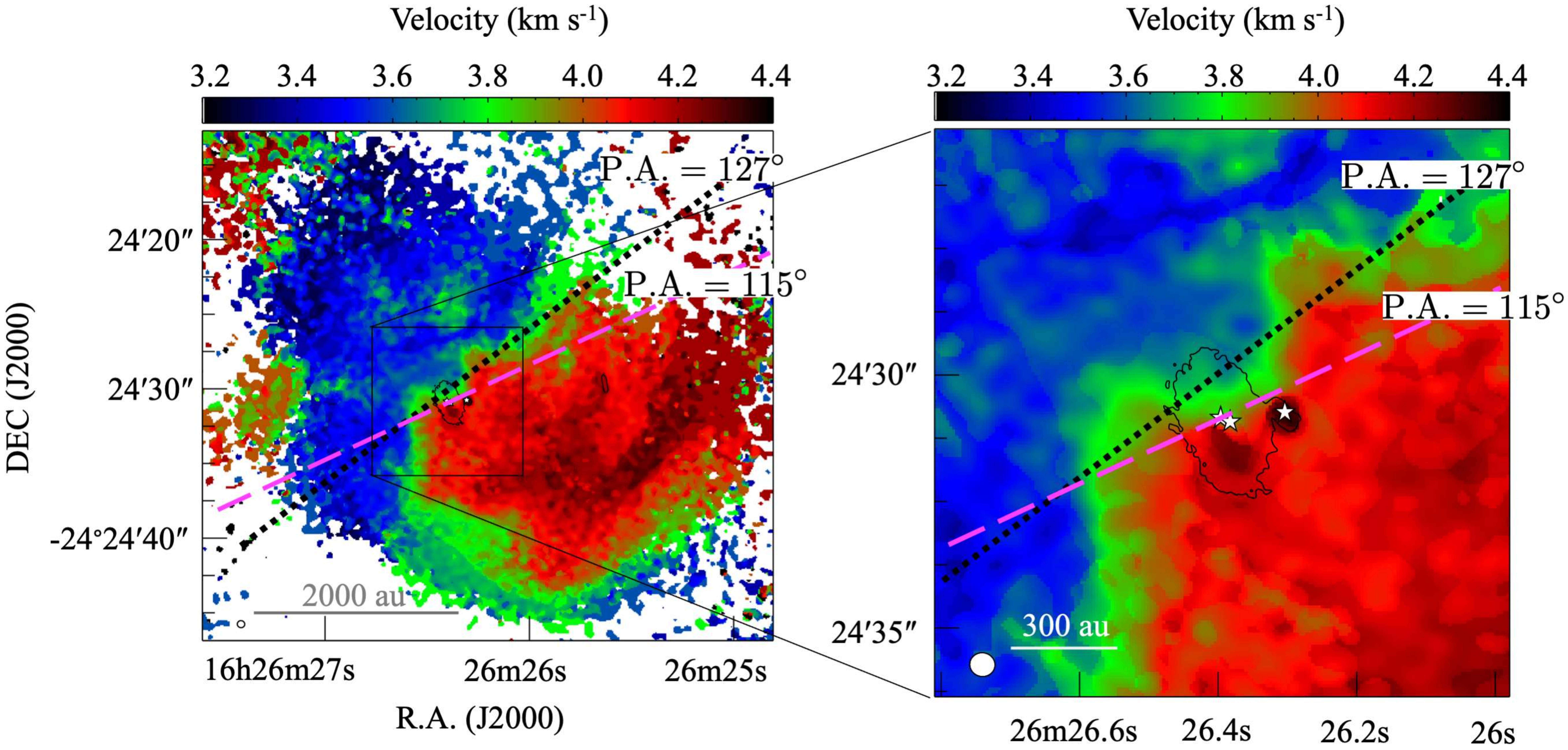}
\caption{Same as Figure \ref{h13cop_mom1} but a velocity  range of  $3.2-4.4$ km s$^{-1}$ with an intensity of $>0.015$ Jy beam$^{-1}$. The right panel is the zoomed-in view around the protostars. Magenta dashed and black dotted lines indicate the rotation axes of the circum-binary disk and envelope, respectively. The beam size is shown in the bottom-left corners.
}
\label{h13cop_mom1_2}
\end{figure*}

The intensity-weighted mean velocity (moment 1) maps of the H$^{13}$CO$^+$ emission are shown in Figure \ref{h13cop_mom1}.
The mean velocity is derived by using a velocity range of $1.2-6.2$ km~s$^{-1}$ with an intensity of $>0.015$ Jy~beam$^{-1}$ (corresponding to $>6\sigma$).
The wide velocity range of $1.2-6.2$ km~s$^{-1}$ includes the high-velocity components of the circum-binary disk rotation in addition to the slow rotation of the envelope.
The overall velocity structure is characterized by blue-shifted emission on the northeast side of the circum-binary disk and red-shifted emission on the southwest side with respect to the systemic velocity of the envelope ($3.8$ km~s$^{-1}$). This velocity gradient in the direction of the large-scale elongation of the H$^{13}$CO$^+$ emission is interpreted as rotation of the envelope.

Interestingly, Figure \ref{h13cop_mom1} seems to show different rotation axes on different scales of the envelope and circum-binary disk because the velocity pattern is twisted from the envelope to the circum-binary disk.
The zoomed-in view of Figure \ref{h13cop_mom1} shows that the velocity gradient in the circum-binary disk is parallel to the major axis of the circum-binary disk, indicating that the rotation motion corresponds to that of the disk.
The magenta dashed lines indicate a direction with a position angle of $115^\circ$, the minor axis of the circum-binary disk, which is derived from Gaussian-fitting \citep{har18}.
The magenta dashed lines agree with the rotation direction of the circum-binary disk, which is tilted with respect to the envelope rotation axis.

To measure the rotation axis of the envelope, we use the mean velocity map made with a velocity range of $3.2-4.4$ km s$^{-1}$, which is narrower than that used in Figure \ref{h13cop_mom1}  because the velocities around the systemic velocity better represent the envelope kinematics without contamination of the rotation of the circum-binary disk.
The mean velocity map of the envelope is shown in Figure \ref{h13cop_mom1_2}.

To measure the rotation axis, we perform a linear fit to the positions whose mean velocity is $3.8\pm0.05$ km s$^{-1}$, assuming that the systemic velocity should be on the rotation axis. 
Furthermore, the fitting area is narrowed to the range from ${+2\farcs6}$ to ${-6\farcs0}$ in R.A. direction and from ${+4\farcs5}$ to ${-4\farcs0}$ in Dec direction from  VLA 1623A1 to avoid the contamination from the infalling components. 
By linear fitting, the rotation axis of the envelope is derived to be a position angle of $\sim127\pm1^\circ$ as shown by the black dotted lines in Figure  \ref{h13cop_mom1_2}.
Even if we select the mean velocity range of $3.8\pm0.2$ km s$^{-1}$, we obtain a similar value of $\sim124\pm1^\circ$.
After these examinations, we conclude that the rotation axis of the envelope is $\sim127\pm4^\circ$.
Figure \ref{h13cop_mom1_2} also plots the rotation axis of the circum-binary disk in the magenta dashed lines.
By comparing these two rotation axes, we confirm that the rotation axes of the circum-binary disk and envelope are misaligned by $\sim12\pm6^\circ$.

Note that the derived rotation axis of the envelope changes slightly if we adopt a different mean velocity range and fitting area. However, the misalignment rotation axes between the envelope and circum-binary disk should be robust because the mean velocity field shows the twisted pattern from the envelope to the circum-binary disk as shown in Figure \ref{h13cop_mom1}.
Such a twisted pattern of the velocity field can also be reproduced by infalling-rotating motion on $\sim100$ au scales \citep[e.g.,][]{yen13,oya16}.
However, the twisted pattern of this source is seen on a larger scale of $\sim1000$ au. Therefore, we suggest that this velocity pattern indicates misalignment.

\begin{figure*}[htbp]
\includegraphics[width=16.cm,bb=0 0 2772 962]{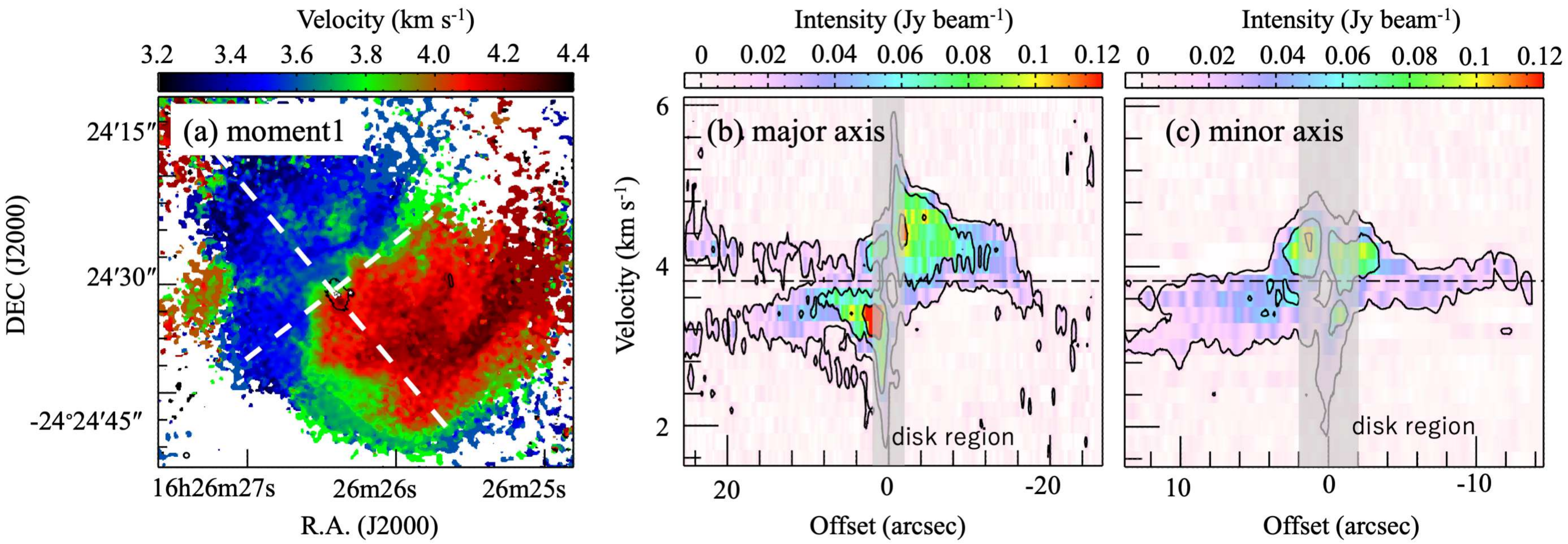}
\caption{Moment 1 map from Figure \ref{h13cop_mom1_2} (a) and PV diagrams of the H$^{13}$CO$^+$ ($J = 3-2$) emission along (b) the major axis and (c) the minor axis.
The shadows shown in the panels of (b) and (c) indicate the circum-binary disk region.
Contours are 0.01, 0.05, and 0.1 Jy beam$^{-1}$. Horizontal dashed lines show the systemic velocity ($V_{\rm sys} = 3.8$ km s$^{-1}$).
}
\label{pv_h13cop}
\end{figure*}

To investigate the kinematics of the envelope, we show the position-velocity (PV) diagrams of the H$^{13}$CO$^+$ emission along with the major and minor axes of the envelope rotation in Figure \ref{pv_h13cop}. 
The PV diagram along the major axis (Figure \ref{pv_h13cop} panel b) shows broad velocity components ($V_{\rm LSR}\sim2-6$ km s$^{-1}$) within $1\arcsec$ from the protostars. These high-velocity components trace the rotation motion of the circum-binary disk. In addition, a velocity gradient on a large scale is found with an offset range from $-20''$ to $20''$. This velocity gradient indicates the rotation motion of the envelope.
The PV diagram along the minor axis (Figure \ref{pv_h13cop} panel c) also contains a high velocity component in the blue shifted emission ($V_{\rm LSR}\sim2-3$ km s$^{-1}$) near the protostars. This high velocity component traces the rotation motion of the circum-binary disk because the minor axis is slightly tilted from the rotation axis of the circum-binary disk.
Contrary to the PV diagram of the major axis, a velocity gradient along the minor axis is not as significant as in the major axis. The emission is almost at the systemic velocity of $3.8$ km s$^{-1}$.

As shown in  Figure \ref{pv_h13cop}, the large velocity width is only seen near the protostars.
Except for this part, the velocity width ($\Delta v$) on the envelope scale is only $\sim0.3-0.5$ km s$^{-1}$. Furthermore, the velocity gradient is seen along the major axis.
Therefore, we suggest that turbulence has only a minor effect on the velocity field of the envelope.

\subsection{ The Outflow Cavity is created by a single outflow or multiple outflows?}\label{subsec:?}

Before investigating the outflow direction and rotation, we discuss the launching source(s) of the CS and CCH outflow cavity.
The integrated-intensity maps of the CS and CCH emission show the simple ``X-shaped'' outflow cavity but the several high-velocity outflows are detected by the CO observations by \citet{san15,har20}.
These CO outflows show wider velocity ranges ($V_{\rm LSR}=-34.5$ to $-0.5$ km s$^{-1}$ and $V_{\rm LSR}=+6.5$ to $31.5$ km s$^{-1}$) than the CS and CCH outflow cavity.
Thus, the CO emission traces the high-velocity outflows, while the CS and CCH emission traces the low-velocity outflow cavity.
\citet{san15} reported that the high-velocity outflows are launched from VLA 1623A and B, and \citet{har20} suggested that there are two outflows launched from A1 and A2.
Thus, the three protostars of VLA 1623A1, A2, and VLA 1623B are suggested to have the high-velocity outflows.

The CS and CCH channel maps (Figures \ref{cs_channel_map} and \ref{cch_channel_map}) indicate that the emission has a velocity gradient from north to south centered around the systemic velocity of 3.8 km s$^{-1}$.
Furthermore, the channel maps of the outflow cavity indicate a symmetric distribution around the systematic velocity of 3.8 km s$^{-1}$.
In particular, the emission distributions around the systematic velocity show the clear ``X-shaped'' structure.
This velocity gradient would be naturally explained by the rotation of a single outflow. 
If material from the circum-binary disk is only accreting onto the circum-stellar disk of VLA 1623A1, the low-velocity outflow would only be launched from the outer edge of the VLA 1623A1 circum-stellar disk.
In the disk wind model, the low- and high-velocity outflows are driven near the disk outer and inner edges, respectively \citep{mac14}.
In contrast, the high-velocity outflows can be launched from the inner edges of the disks due to the accretion toward the protostars.
The low-velocity outflow has a wider opening angle than the high-velocity outflows. Therefore, the CO high-velocity outflows are encompassed by the CS and CCH emission, which is consistent with the recent MHD simulation of a binary formation \citep{saiki20}.

The other possibility is that the two independent outflows may produce such velocity gradient of the cavity structure.
As similar to the rotation scenario, \citet{har20} showed that the blue shifted outflows are located in the north, and the red shifted outflows are located in the south (Figure \ref{cs_channel_map}).
These high-velocity outflows are encompassed by the CS and CCH outflow cavity.
Therefore, it would be possible that the velocity gradient of the outflow cavity is produced by the entrainment of these high-velocity outflows.

Our current observations cannot conclude whether the velocity gradient of the outflow is caused by rotation or the entrainment of the high-velocity outflows.
However, we discuss the outflow properties such as the outflow axis, angular momentum, and launching radius by assuming the outflow rotation scenario in the discussion section (Section \ref{subsec:outflow}).

Note that the observed velocity gradient would not be explained by the precession motion of the binary protostars because a ``S-shaped'' wiggling structure is not identified in the channel maps.
If the outflow has a precession motion and produces the observed velocity gradient, a curved trajectory is expected \citep[e.g.,][]{fen98}.
However, the channel maps indicate that the outflow cavity structure runs from north to south without any wiggling structures and with increasing velocity.

To find the launching position in detail, we reanalyzed the CS data by CASA task {\it clean} with a robust parameter of $-2$.
The image is obtained with a slightly better spatial resolution of $0\farcs474\times0\farcs435$.
Figure \ref{cs_channel_map_zoom} shows the zoomed-in views of the CS channel maps around the protostars with a better spatial resolution.
The velocity gradient can be seen in the vicinity of the protostars.
Even though the spatial resolution of the CS emission is not enough to distinguish the source A1 and A2, the outflow seems to be launched in the vicinity of source A1.
This can be seen in the channels at $V_{\rm LSR}=2.4$, 4.4, and 4.8 km s$^{-1}$.
For example, the outflow cavity structures seen at $V_{\rm LSR}=2.4$ and 4.8 km s$^{-1}$ cross at the position of A1 (see the dashed lines in the channel maps of $V_{\rm LSR}=2.4$ and  4.8 km s$^{-1}$).
However, our spatial resolution is not high enough to determine the launching source definitively.
Further investigations with higher spatial resolution are needed to identify the launching position(s) of the CS and CCH outflow cavity and distinguish the outflow rotation from the entrainment of the high-velocity outflows.

\begin{figure*}[htbp]
\includegraphics[width=16.cm,bb=0 0 2222 1282]{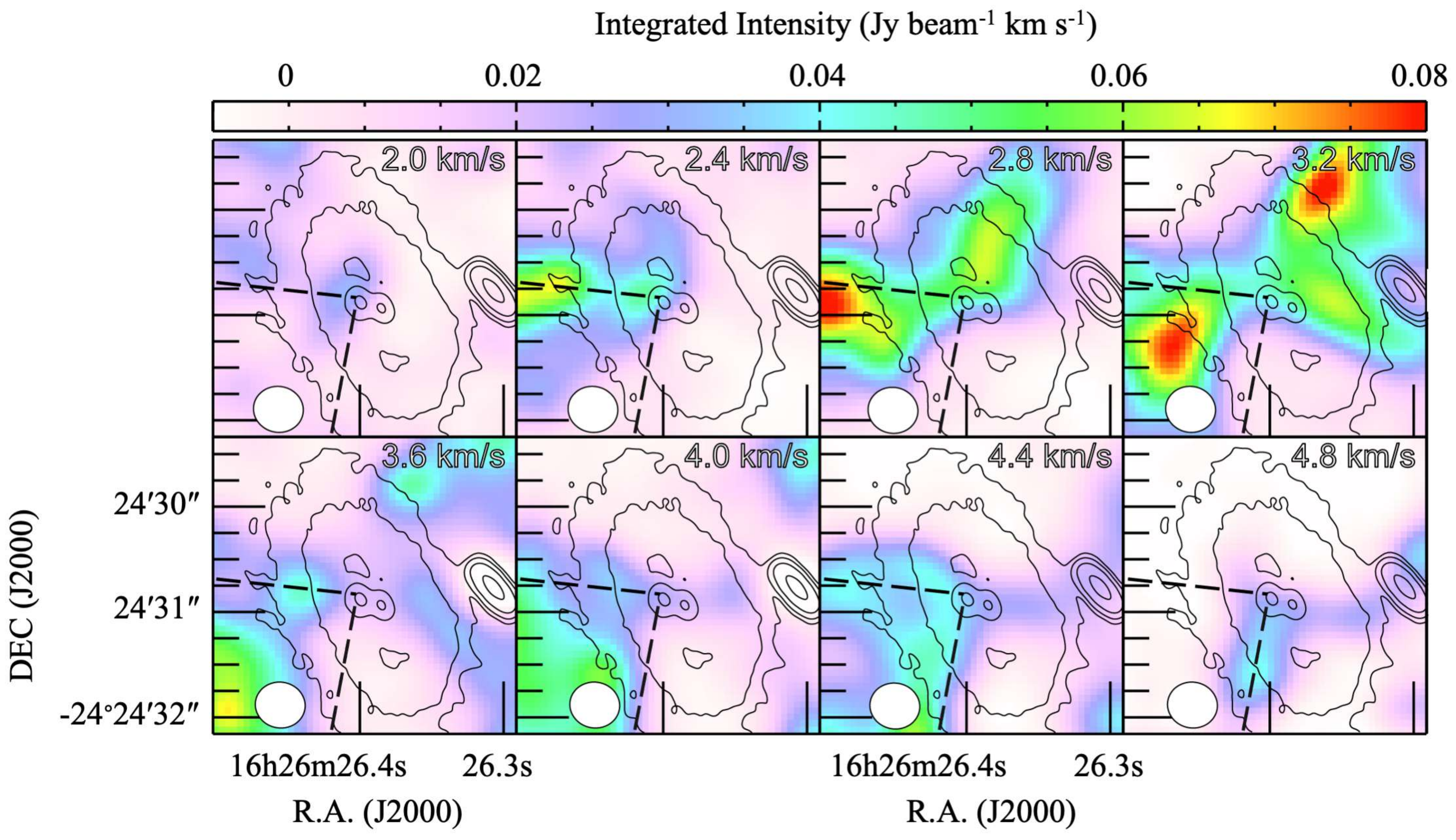}
\caption{Same channel maps as Figure \ref{cs_channel_map} but zoomed-in views around the protostars. The contour levels are the same as in Figure \ref{cont}. The channel width and step are 0.4 km s$^{-1}$, and the channel centroid velocities are labeled at the upper right corner of each panel. The dashed lines indicate the outflow cavity structures seen at the channel maps of $V_{\rm LSR}=2.4$ and $4.8$ km s$^{-1}$.
The beam size is shown in the bottom-left corners.
}
\label{cs_channel_map_zoom}
\end{figure*}

\subsection{Outflow Axis, Angular Momentum, and Launching Radius}\label{subsec:outflow}

 In the previous subsection, we discussed the outflow launching source.
From the channel maps of the CS and CCH emission, we suggested a possibility that the velocity gradient is caused by the outflow rotation launched from VLA 1623A1.
Based on this assumption, we investigate the outflow properties of the axis, angular momentum, and launching radius.
We note that it may also be possible that the CS and CCH outflow cavity is caused by the entrainment of the two high-velocity outflows rather than rotation.
In that case, the discussion of this subsection (Section \ref{subsec:outflow}) would need to be re-investigated.

\subsubsection{Outflow Axis}\label{subsubsec:outflow_axis}

\begin{figure*}[htbp]
\includegraphics[width=16.cm,bb=0 0 2447 1398]{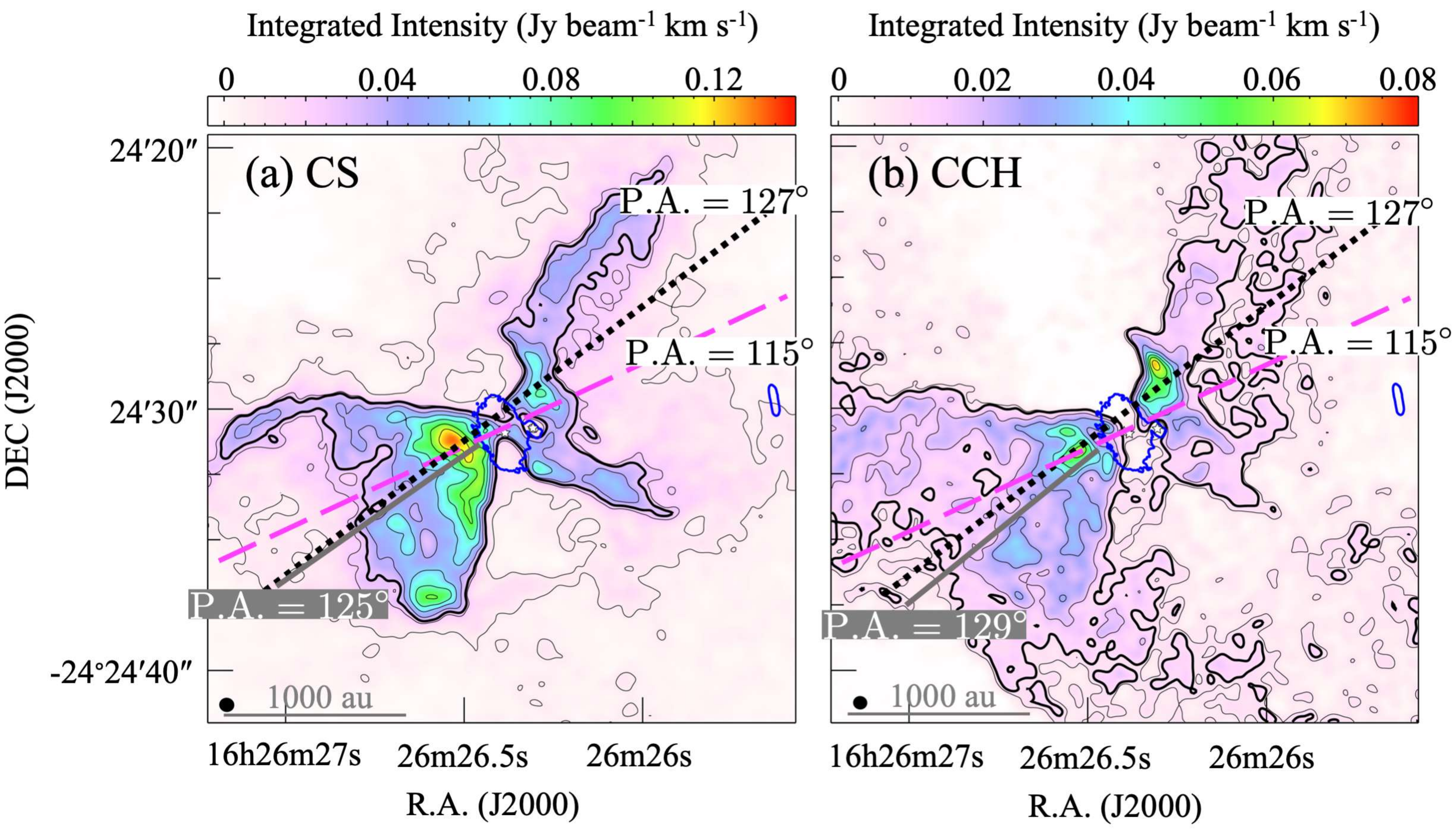}
\caption{Integrated intensity maps of the (a) CS and (b) CCH emission overlaid with the 0.87 mm dust continuum emission \citep{har18} in blue thick contours. The velocity  range for integration is $3.4-4.2$ km s$^{-1}$. Thin black contours of the CS emission start from $5\sigma$ and have intervals of $15\sigma$, where $1\sigma=1.2$ mJy beam$^{-1}$ km s$^{-1}$. Thin black contours of the CCH emission start from $5\sigma$ and have intervals of $7\sigma$, where $1\sigma=1.4$ mJy beam$^{-1}$ km s$^{-1}$. Thick black contours of the CS and CCH emission encompass the outflow cavity structure and are $25\sigma$ and $7\sigma$, respectively. The 0.87 mm dust continuum contours are at a level of 0.93 mJy beam$^{-1}$.
The magenta dashed, black dotted, and grey lines indicate the minor axis of the circum-binary disk, rotation axis of the envelope, and outflow direction, respectively. The beam size is shown in the bottom-left corners.
}
\label{outflow_direction}
\end{figure*}

To measure the outflow direction, we use the integrated intensity maps of the CS and CCH emission shown in Figure \ref{outflow_direction}.
The velocity range for the integration is selected to be $V_{\rm LSR}=3.4-4.2$ km s$^{-1}$ around the systemic velocity of the outflow ($\sim3.8$ km s$^{-1}$).

The thick black contours in Figure \ref{outflow_direction} show the $25\sigma$ and $7\sigma$ noise levels of the CS and CCH integrated intensities, respectively. 
These contours encompass the outflow cavity structure.
First, we derive the directions of the outflow cavity edges (north and south) by performing the linear fitting of these contours.
Then, we derive the outflow direction by averaging the position angles of these two edges.

We focus only on the southeast lobe, because the northwest outflow is affected by absorption against the VLA 1623B disk.
In addition, the fitting area is narrowed to within $6.2\arcsec$ (corresponding to 850 au) from source A1 because the contours ($25\sigma$ of CS and $7\sigma$ of CCH) do not follow the outflow structure outside of this area.

By fitting the contours, the position angles of the outflow cavity edges (north and south directions) are derived to be $\sim84\pm0.4^\circ$ and $\sim166\pm1^\circ$ for the CS outflow, and $\sim91\pm0.5^\circ$ and $\sim167\pm1^\circ$ for the CCH outflow.
Therefore, we derive the outflow directions of $\sim125\pm1^\circ$ for the CS outflow and $\sim129\pm1^\circ$ for the CCH outflow. These outflow directions are close to each other.
Furthermore, the outflow direction of $\sim125-129^\circ$ is also consistent with that of the CO high velocity outflow of $128^\circ$ (outflow-2 launched from A1 or A2) measured by \citet{har20}.
We note that the opening angle of the CS outflow cavity looks slightly wider than the CCH outflow cavity since the CS emission traces not only the outflow cavity structure but also the accretion flows. It also causes the slight difference of the outflow directions between the CS and CCH emission.

The outflow directions are shown in the thick grey lines in Figure \ref{outflow_direction}.
We also plot the rotation axis of the envelope and minor axis of the circum-binary disk in the magenta dashed and black dotted lines, respectively.
Note that the minor axis of the circum-binary disk can be regarded as the rotation axis of the circum-binary disk.
These directions indicate that the rotation axis of the envelope is consistent with the outflow direction, while the rotation axis of the circum-binary disk is slightly different from the others. 
We discuss possible origins of these different rotation axes in Section \ref{sec:mis}.

\subsubsection{ Rotation and Launching Radius of the Outflow}\label{subsubsec:outflow_rot}

\begin{figure*}[htbp]
\includegraphics[width=16.cm,bb=0 0 2418 1411]{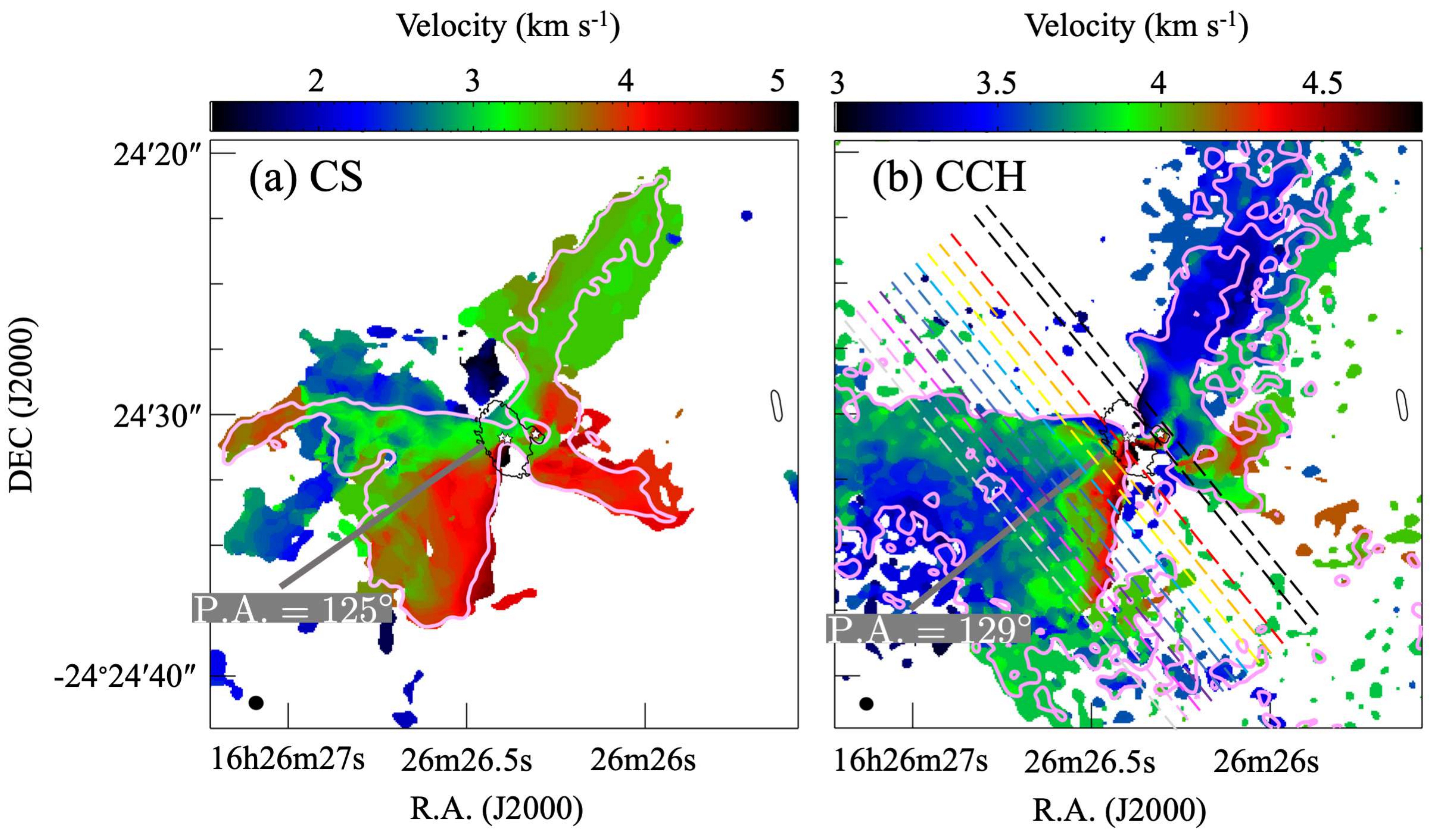}
\caption{Mean velocity (moment 1) maps of the CS (a) and CCH (b) emission. The mean velocity of the CS emission is derived by using a velocity range of $0.6-5.4$ km  s$^{-1}$ with an intensity of $>0.05$ Jy beam$^{-1}$, while that of the CCH emission is derived by using a velocity range of $2.6-6.2$km s$^{-1}$ with an intensity of $>0.02$ Jy beam$^{-1}$ to avoid the contamination from the hyperfine emission.
Magenta contours indicate $25\sigma$ and $7\sigma$ of the CS and CCH integrated-intensity emission, respectively.
Black lines indicate the outflow directions.
The dashed lines perpendicular to the outflow axis indicate the cuts for the position-velocity diagrams shown in Figures \ref{outflow_roation_pv} and \ref{outflow_rotation}.
The colors of the cuts across the eastern lobe are used in Figure \ref{outflow_rotation}. (b) Same as
panel (a), but for the CCH emission. The beam size is shown in the bottom-left corners.
}
\label{outflow_direction_mom1}
\end{figure*}

In this subsection, we investigate outflow properties by analyzing the angular momentum and the launching radius of the outflow.
The mean velocity maps of the CS and CCH emission are shown in Figure \ref{outflow_direction_mom1}.
We confirm that the velocity gradient is found to be perpendicular to the outflow direction. The north part of the outflow is blue shifted, while the south part is red shifted.
 We suggested that this velocity gradient indicates the outflow rotation in Section \ref{subsec:?}.
However, we should note that there might be the possibility that the velocity gradient is produced by the two independent outflows.

The magenta contours indicate the $25\sigma$ and  $7\sigma$ noise levels of the CS and CCH emission, respectively, encompassing the outflow cavity structure. The velocity field of the CCH emission shows a gradient within this contour.
In contrast, the velocity field of the CS emission shows the gradient even outside the $25\sigma$ contour.
This may indicate that the velocity gradient shown in the CS emission may be contaminated by the accretion flows or envelope materials because the velocity gradient in the CS is in the same direction as the rotation of the envelope.
Therefore, we focus only on the CCH emission in the following analysis.

\begin{figure*}[htbp]
\includegraphics[width=18.cm,bb=0 0 1218 1373]{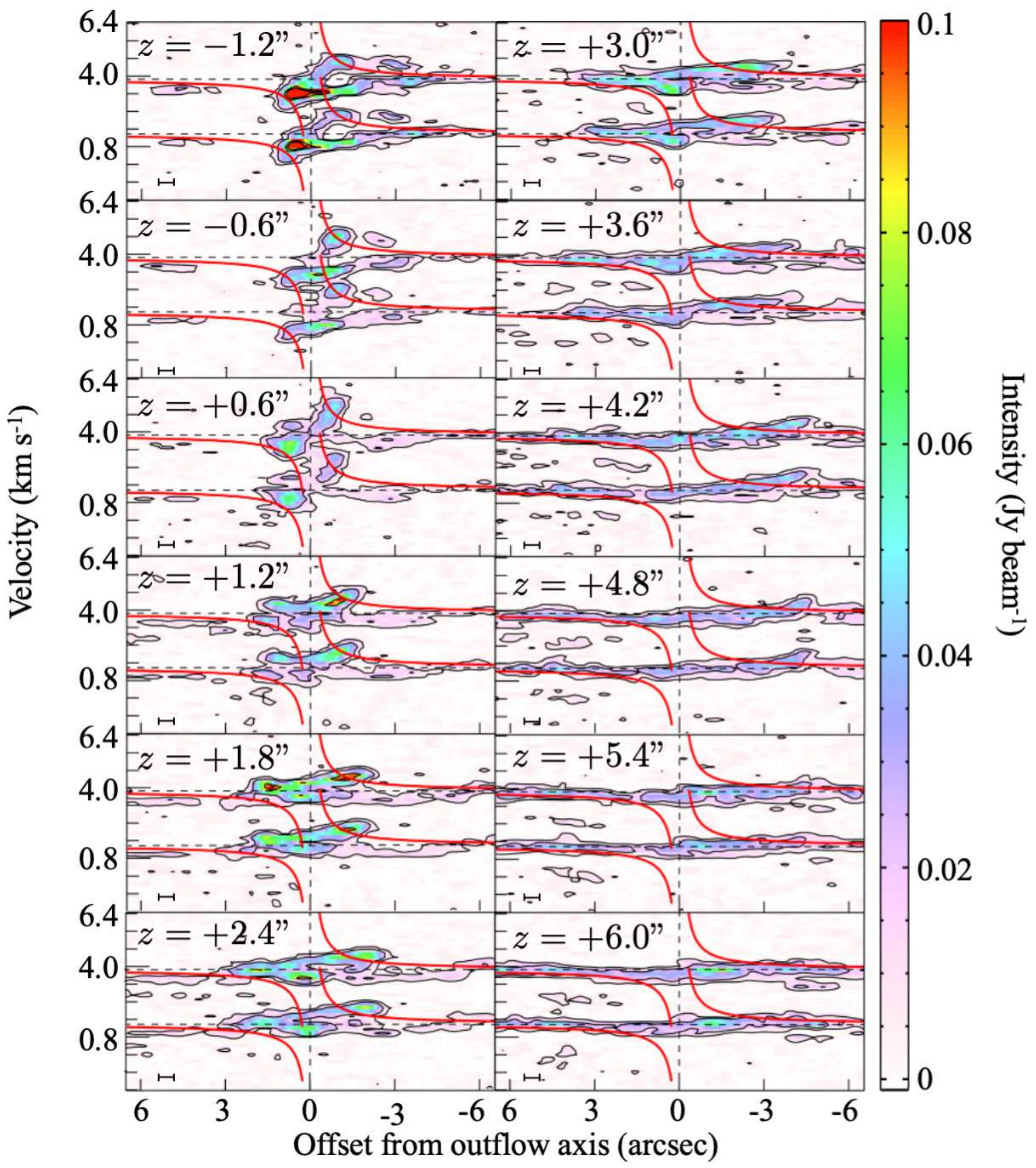}
\caption{Position-velocity diagrams of the CCH ($N=3-2$, $J=5/2-3/2$) emission along the lines perpendicular to the outflow axis (shown in Figure \ref{outflow_direction_mom1}). The distances of these lines to the central source are labeled at the upper-left corners of each panel ($z>0$ for the southeast lobe and $z<0$ for the northwest lobe). The contours are 6, 20, 80, and 100 mJy beam$^{-1}$. The rest frequency of the $F=4-3$ hyperfine line is used, so that the $F=3-2$ hyperfine line appears at a blue-shifted velocity. The red curves correspond to a constant angular momentum of 100 au km s$^{-1}$ (see Section \ref{subsubsec:outflow_rot}). The black bar at the lower-right corner of each panel indicates the beam size.
}
\label{outflow_roation_pv}
\end{figure*}

To investigate the rotation of the outflow, we analyze the CCH emission in a similar way to \citet{zha18}.
We show the PV diagrams of the CCH emission in the southeast lobe and the base of the northwest lobe, along the lines perpendicular to the outflow axis (the lines are shown in Figure \ref{outflow_direction_mom1}).
 We set the center at the outflow axis, and the negative value is the northern part of the outflow cavity, while the positive value is the southern part of the outflow cavity.
The outflow axis ($z$) passes through the  VLA 1623A1 protostar and has a position angle of $129^\circ$.
The lines of the PV diagrams are taken  for every $0\farcs6$ from $z=-1\farcs2$ to $+6\farcs0$ except for $z=0$.

The PV diagrams in Figure \ref{outflow_roation_pv} indicate that the higher velocity components are found closer to the protostar, and the emission has a wider spatial distribution at farther distances from the protostar.
These features are consistent with the outflow from NGC 1333 IRAS 4C \citep{zha18}.
A red curve shown in each panel indicates a constant angular momentum of 100 au km s$^{-1}$ with respect to the outflow axis.
The CCH emission seems to follow this curve, suggesting constant angular momentum of the outflow.
Note that the PV diagrams with $z\geq2\farcs4$ do not show any sign of higher velocity components around the outflow axis position (offset $\sim0\farcs0$). This is because the CCH emission traces the outflow cavity wall. Even if the CCH emission is observed apparently on the outflow axis, the emission region is the cavity wall at the near and far sides of the outflow axis. The slow rotation velocity of the cavity wall is due to a relatively large linear distance from the outflow axis.
The median velocity of $3.8$ km s$^{-1}$ (the horizontal dashed lines) does not change with $z$ from the systemic velocity of the envelope, indicating a nearly edge-on configuration.
We suggest that the CCH emission traces the walls of the outflow cavity instead of inner part of the outflow. Thus, high velocity components at large distances from the central source cannot be seen by the CCH emission.

\begin{figure}[htbp]
\includegraphics[width=8.cm,bb=0 0 1716 1538]{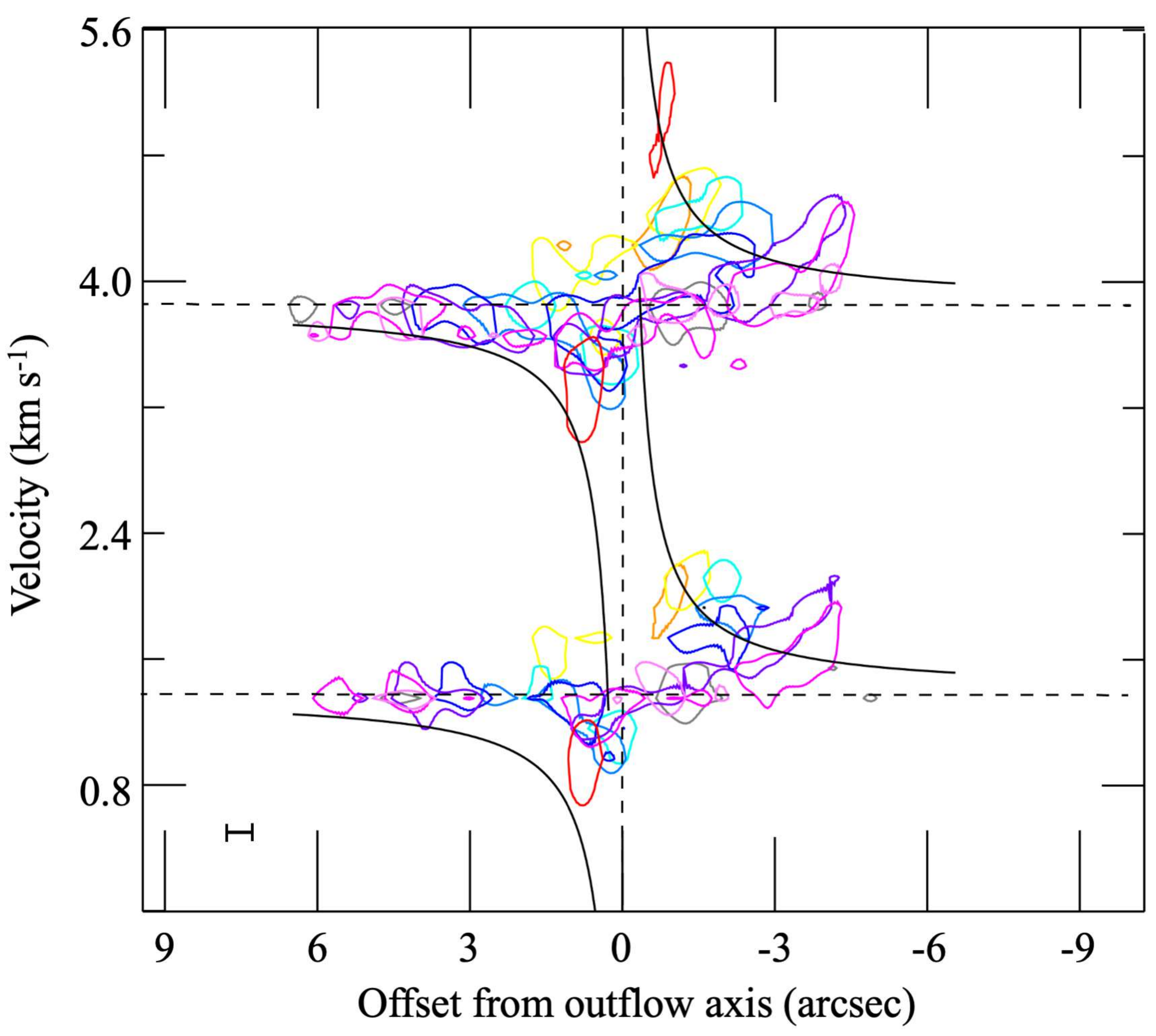}
\caption{Position-velocity diagram showing the emission peaks of different panels of the southeast lobe in Figure \ref{outflow_roation_pv} in one panel. The color corresponds to the different cuts shown in Figure \ref{outflow_direction_mom1}. The contours are at levels of 0.6 times the maximum intensity in each panel. The curves are the same as those in Figure \ref{outflow_roation_pv}.
}
\label{outflow_rotation}
\end{figure}

To show the outflow rotation more clearly, the contours of the PV diagrams for the southeast lobe ($z\geq0\farcs6$) are overlaid to form the composite PV diagram shown in Figure \ref{outflow_rotation}. 
The colors depict the position of the cut lines shown in Figure \ref{outflow_direction_mom1}.
The contours are set to be $0.6\times I_{\rm peak}$, where $I_{\rm peak}$ is the maximum intensity in each panel.
The contours follow the expected curve for constant angular momentum of 100 au km s$^{-1}$.
Interestingly, the angular momentum of 100 au km s$^{-1}$ is quite similar to that for the NGC 1333 IRAS 4C outflow \citep{zha18}.
Because both VLA 1623A1 and NGC 1333 IRAS 4C are low-mass ($\sim0.2$ $M_\odot$) protostars in the Class 0 stage, it is not surprising that these sources have a similar outflow properties.
However, the rotation feature on the PV diagrams in this outflow is not as clear as that of NGC 1333 IRAS 4C. Further investigation with better spatial resolution observations will allow us to reveal the rotation motion more clearly.

Next, we estimate the launching radius of the outflow from the angular momentum.
Following the method described by \citet{and03}, the wind-launching radius can be constrained by the outflow rotation, the poloidal velocity of the outflow ($v_{p}$), and the protostellar mass.
Because the outflow is aligned with the plane of sky, the poloidal velocity is difficult to measure. Therefore, we assume a typical value of $v_{p}=3-10$ km~s$^{-1}$, as is the case for NGC 1333 IRAS 4C. We also assume that the  poloidal velocity of the outflow is much higher than the toroidal (rotation) velocity.
The total mass of  VLA 1623A1 and A2 is assumed to be $0.3-0.5$ $M_\odot$, and VLA 1623A has been considered to be an equal-mass binary system \citep{har18,kaw18}.
Therefore, we simply assume the half value of $0.2$ $M_\odot$ for the  VLA 1623A1 protostellar mass.
With these assumptions, the launching radius is derived to be $5-16$ au.
Even if we use a higher mass of $0.4$ $M_\odot$ assuming the launching point in the circum-binary disk, the launching radius is still $5-10$ au.
This value is consistent with the fact that the launching position is not resolved by our observations.
Since the separation of the binary is 28 au, we conclude that the outflow is launched from the circum-stellar disk of VLA 1623A1 (or may be A2) rather than the circum-binary disk.

The derived launching radius ($5-16$ au) is comparable to other outflow launching radii such as TMC1-A \citep{bje16}, Orion Source I \citep{hir17},  HH 212 \citep{tab17,lee18}, HH 46/47 \citep{zha16}, NGC 1333 IRAS 4C \citep{zha18}, and Monoceros R2-IRS2 \citep{jim20}.  
The rotation in the inner jet, on the other hand, has been reported in some other sources with a launching radius of $<0.1$ au such as B335 \citep{bje19} and HH 212 \citep{lee17}.
Theoretical models predict the launching of fast jet and slow outflow at the inner disk edge and outer disk radius, respectively \citep{mac14}.
If this is the case, our results are consistent with a wide-angle disk wind model launched from relatively large radii in the disk.

\begin{figure*}[htbp]
\includegraphics[width=16.cm,bb=0 0 1916 1596]{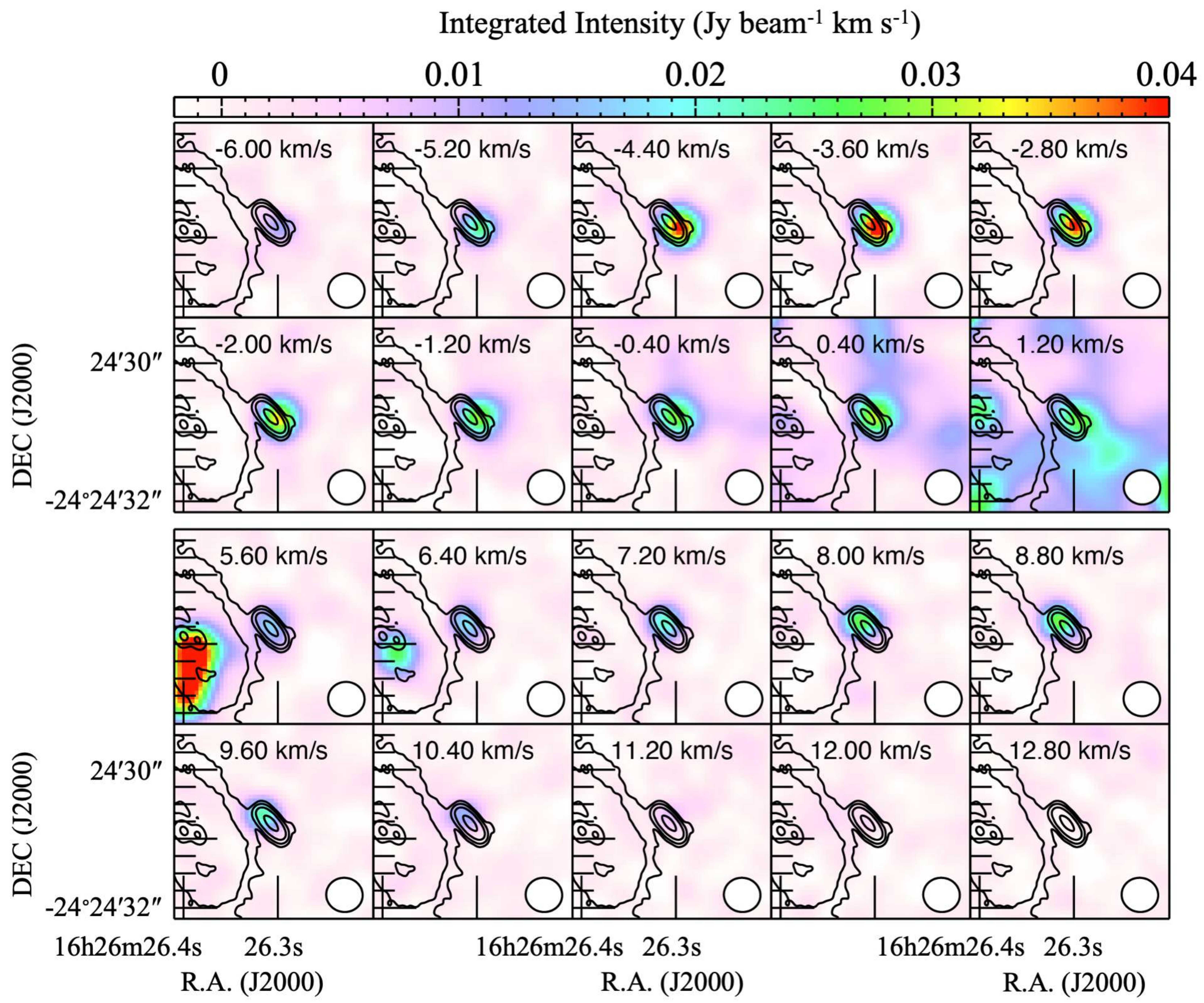}
\caption{Channel maps of the  CS emission from the high velocity components, overlaid with the 0.87 mm dust continuum emission \citep{har18} in black contours. The channel width and step are 0.8 km s$^{-1}$, and the channel centroid velocities are labeled at the upper center of each panel. The beam size is shown in the bottom-right corners.
}
\label{disb_rotation_channel_map}
\end{figure*}

\begin{figure*}[htbp]
\includegraphics[width=16.cm,bb=0 0 2999 960]{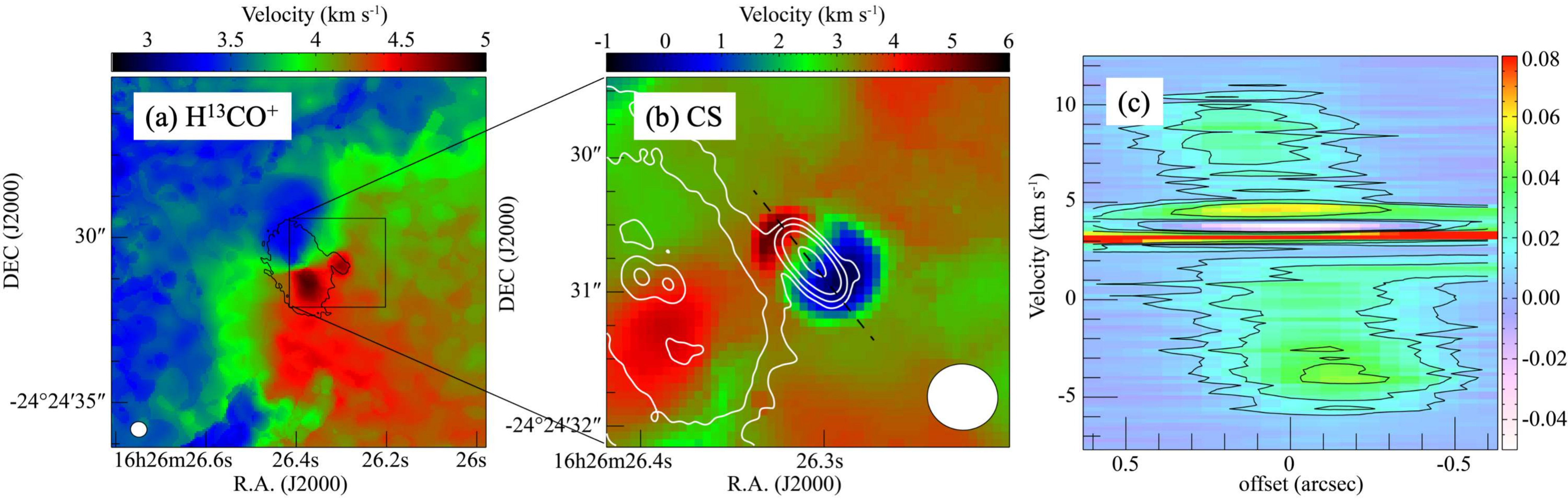}
\caption{(Panel a) Intensity-weighted mean velocity (moment 1) map of the H$^{13}$CO$^+$ emission from Figure \ref{h13cop_mom1}.
(Panel b)  Mean velocity (moment 1) map of the CS emission toward the VLA 1623B disk. The mean velocity is derived using a velocity range of $-8$ to $15$ km s$^{-1}$ with an intensity of $>0.01$ Jy beam$^{-1}$. White contours indicate the 0.87 mm dust continuum emission.
(Panel c) PV diagram of the CS emission toward the circum-stellar disk of VLA  1623B. The direction is indicated as the black dashed line in the panel (b). The contours are 0.01, 0.02, and 0.04 Jy beam$^{-1}$, respectively.
}
\label{diskb_rotation_pv}
\end{figure*}

\subsection{The reversed velocity gradient of the disk around VLA 1623B}\label{subsec:rot_diskb}

 Here, we report the discovery of the velocity gradient of the VLA 1623B disk, which has the opposite sense to the others of the envelope, circum-binary disk, and the outflow.

As shown in the integrated intensity map of the CS emission (Figure \ref{cs_cch_mom0}), the peak emission is found toward the VLA 1623B disk due to the contribution of the high velocity component.
Figure \ref{disb_rotation_channel_map} shows channel maps of the blue- and red-shifted high-velocity components of the CS emission.
The velocity range for the integration and velocity step of the channels are 0.8 km s$^{-1}$. The channel maps of the blue-shifted component show the velocity ranging from $-6.4$ to $1.6$ km s$^{-1}$, while the channel maps of the red-shifted component show the velocity ranging from $5.2$ to $13.2$ km s$^{-1}$.
We confirm that the high velocity emission is mainly detected in the VLA 1623B disk.

From the channel maps of Figure \ref{disb_rotation_channel_map}, we find a velocity gradient along the disk major axis.
The blue-shifted emission is detected in the southern part of the disk, while the red-shifted emission is detected in the northern part, indicating the rotation motion.
Interestingly, this velocity gradient is in the opposite sense to that of the envelope, outflow, and VLA 1623A circum-binary disk.

Figure \ref{diskb_rotation_pv} shows the mean velocity maps of the H$^{13}$CO$^+$ and CS emission overlaid on the 0.87 mm dust continuum image.
The mean velocity map of the H$^{13}$CO$^+$ emission is the same as that of Figure \ref{h13cop_mom1} which is used for revealing  the rotation motion of the circum-binary disk. 
We confirm that Figure \ref{diskb_rotation_pv} shows the  velocity gradient of the VLA 1623B is reverse to that of the circum-binary disk.
The reversed velocity gradient is also confirmed by the PV diagram along the disk major axis shown in the right panel (c).

Here, we roughly estimate the dynamical mass of VLA 1623B from the velocity gradient.
The dynamical mass ($M_{\rm dyn}$) is calculated as 
\begin{equation}
M_{\rm dyn}=\frac{r v^2_{\rm rot}}{G}\sim0.01 \left(\frac{r}{10 \ \rm au}\right)\left(\frac{v_{\rm rot}}{1\ \rm km s^{-1}}\right)^2~M_{\odot},
\end{equation}
where $r$ and $v_{\rm rot}$ are the radius and rotation velocity of the disk, respectively.
The disk is assumed to be edge-on.
As shown in Figure \ref{diskb_rotation_pv}, the velocity gradient of the rotation is roughly $\sim39$ km s$^{-1}$ arcsec$^{-1}$ ($V_{\rm LSR}=-5$ to $9$ km s$^{-1}$ within a $0\farcs35$ distance).  Although the disk structure is not resolved by our observation, we roughly assume a disk radius of $0\farcs2$ (corresponding to 27 au) because the peak distance between the blue and red components is about $\sim0\farcs4$.
Then, the dynamical mass ($M_{\rm dyn}$) is derived to be $\sim1.7$ $M_{\odot}$, which is much larger than the total mass of the VLA 1623A ($\sim0.4$ $M_{\odot}$).
If the CS-emitting region is smaller than the above estimate, the derived mass may be smaller: it would be $0.2 - 0.4$ $M_{\odot}$ if the radius of the emitting region were $13 - 17$ au (corresponding to $\sim0\farcs1$). 
Higher spatial resolution observations are needed to derive a more precise value for the dynamical mass of VLA 1623B.

It is very surprising that the velocity gradient of the VLA 1623B disk is completely opposite to the other velocity gradients of the envelope, circum-binary disk, and outflow.
The position angle of the VLA 1623B disk is derived to be $41.4\pm0.5^\circ$ from Gaussian fitting of the dust continuum image \citep{har18}.
Thus, the rotation axis corresponds to $131.4\pm0.5^\circ$, which is similar to the outflow axis ($\sim129^\circ$) and the envelope rotation axis ($\sim127^\circ$).
We discuss the origin of the reversed velocity gradient of the VLA 1623B disk in the next section.

Note that opposite rotations in close binary disks have recently been identified in a high-mass star forming region \citep{tan20}, suggesting that these different rotations in close binary systems might be a common occurrence and are established in the early phases of star formation.

Although previously it has been proposed that VLA 1623B corresponds to a shocked cloudlet \citep{mau12,har20}, the clear rotation pattern in these observations further confirms that the structure corresponds to a protostellar disk. 

\section{The origin of the misalingment} \label{sec:mis}

In the previous sections, we found that the rotation axes of the envelope and low-velocity outflow are consistent, while the rotation axis of the circum-binary disk (VLA 1623A) is slightly tilted by $\sim12\pm6^\circ$.
Even though the velocity gradient of the low-velocity outflow will be explained either by rotation or entrainment of the two high-velocity outflows, the direction of the low-velocity outflow is approximately consistent with the envelope rotation axis as shown in Section \ref{subsubsec:outflow_axis}.
In addition,  the velocity gradient of the VLA 1623B disk is opposite to the other velocity gradients.
The several high-velocity outflows are reported by the CO observations, which would be launched from the circum-stellar disks around the VLA 1623A1, A2, and VLA 1623B protostars \citep{san15,har20}.
In contrast, we suggest that the low-velocity outflow may be launched from the outer edge of the VLA 1623A1 disk at a radius of $r\sim5-16$ au.
These high-velocity outflows and single low-velocity outflow are well aligned with the envelope rotation.

According to these results, Figure \ref{view} illustrates a schematic view of the possible rotation motions on various spatial scales, and Table \ref{table2} lists the derived position angles of the rotations and disk inclinations.

To identify the origin of the misalignment of the circum-binary disk of VL1623A, the magnetic field direction is expected to be an important parameter.
Recent ALMA dust continuum polarization observations detected the polarization in the circum-binary disk and the VLA 1623B disk \citep{sad18,har18}.
\citet{sad18} found that the polarization morphology is well explained by a static, oblate spheroid model with a poloidal magnetic field, assuming that the dust polarization traces the magnetic fields in the disk, similar to other disks \citep{alv18,oha18}.
Furthermore, the direction of the poloidal magnetic field is consistent with the circum-binary disk minor axis.
Thus, the direction of the global magnetic field is misaligned with the rotation axis of the envelope and outflow by $\sim12^\circ$ (Figure \ref{view}).

Based on our results together with the previous observations, the three dimensional structures of the disks are revealed.
We note, however, that these disk inclinations are based on an interpretation of the velocity structures as purely due to outflow rotation.
If the multiple protostars and multiple outflows create complex velocity structures, the three dimensional structures shown in Figure \ref{view} would be changed. 

\begin{table*}[ht]
\caption{ List of Derived Position Angles at Various Scales}
\scalebox{1.0}{
\begin{tabular}{lcccc}
\hline \hline
 Region &Probed by & Position Angle & Inclination & Scale \\
 \hline
Envelope & H$^{13}$CO$^+$& $127\pm4^\circ$ & & $\sim2000$ au \\   
Outflow & CS & $125\pm1^\circ$ & & $\sim2000$ au \\   
Outflow & CCH & $129\pm1^\circ$ & & $\sim2000$ au \\   
circum-binary disk$^a$  & 0.87 mm dust continuum & $115\pm4^\circ$$^c$ & $\sim55^\circ$$^c$ & $\sim300$ au \\   
 (VLA 1623A) & &\\
circum-stellar disk$^b$  & 0.87 mm dust continuum &  $131.4\pm0.5^\circ$$^c$ & $\sim74^\circ$$^c$ & $\sim100$ au \\   
 (VLA 1623B) & &\\
\hline
\label{table2}
\end{tabular}}
\begin{flushleft}
\tablecomments{\\
$^a$ The rotation is seen in H$^{13}$CO$^+$.\\
 $^b$ The rotation is seen in CS. \\
 $^c$ The position angle and inclination are derived by \citet{har18}.}
\end{flushleft}
\end{table*}

\begin{figure*}[htbp]
\includegraphics[width=15.cm,bb=0 0 2798 1674]{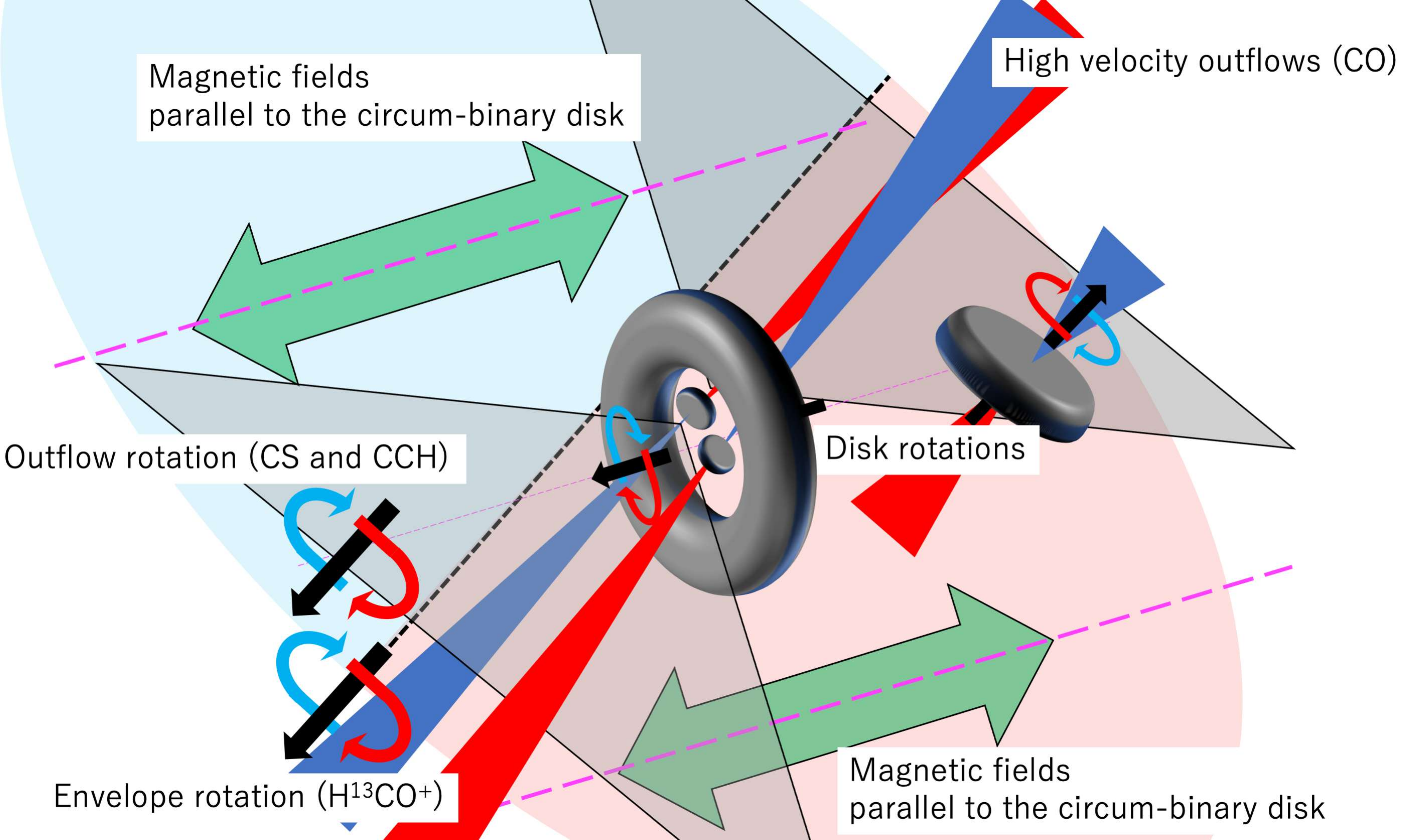}
\caption{A schematic view of the various rotation motions in the VLA 1623 region studied in this work.
}
\label{view}
\end{figure*}

\subsection{The origin of the misalignment of the circum-binary disk}\label{subsec:mis_cb}

The VLA 1623 region has complicated structures such as the multiple system. 
Our observations found that the circum-binary disk is misaligned with the envelope rotation.
It may be possible that the binary motion or the third stellar component (VLA 1623B) perturbs the circum-binary disk.
In this case, it is expected that the circum-stellar disks and outflow direction are also perturbed by such orbital motions. 
The change of the outflow direction is recently reported by \citet{oko21} in another young stellar object, IRAS 15398$-$3359.
However, \citet{har20} estimate a precession amplitude of $0.39\pm0.02^\circ$ for the outflow, which is much smaller than the misalignment of $\sim12^\circ$.

The misalignment of a disk from an envelope has been investigated with numerical simulations by taking into account the magnetic field, turbulence, and rotation of a dense core \citep[e.g.,][]{mat04,hen09,li13,sei13}.
Recent hydro-dynamical simulations suggest that a circum-binary disk and circumstellar disks around individual stars can be misaligned with each other in turbulent cores because the angular momentum of accreting material changes with evolution \citep{bat18}.
Thus, the orientations of the circum-binary disk and circumstellar disks could be more chaotic.
Such misalignment between a circum-binary disk and circum-stellar disks has recently been found by \citet{alv19,mau20}.

However, this scenario seems to be inconsistent with the well-aligned structures of the envelope and outflow in this source.
The same rotation axes of the envelope and outflow suggests that the angular momentum of the accreting material does not change significantly with evolution.
Furthermore, the turbulent motion does not appear to affect the velocity fields on scales of the envelope, as shown in Figure \ref{pv_h13cop}.

Non-ideal MHD simulations have also investigated the outflow and disk structures in relation to the angular momentum and magnetic field directions \citep{mat17,hir19}. 
\citet{hir19} investigated the relationship between outflows, rotation axes, and magnetic field directions in dense cores. They showed that low-density collapsing gas is tightly coupled to the magnetic field, resulting in the misalignment of the outer region of the disk with the rotation axis. In addition, they indicated that the inner part of the disk is well aligned with the core rotation axis because the rotation or centrifugal forces overcome the magnetic effects, due to magnetic dissipation in high-density regions.
The degree of the misalignment is determined by the initial conditions.
The misalignment of $\sim12^\circ$ found by our study is comparable with these simulations.
Thus, a misalignment or warped disk can be formed due to the different directions of the rotation axis and magnetic fields.

This is indeed the case for VLA 1623 as shown in Figure \ref{view}.
The circum-stellar disks of the VLA 1623A1 and A2 protostars may be formed in high-density inner region, where the magnetic field is dissipated, and these circum-stellar disks are aligned with the rotation axis.
Because the dissipation of the magnetic field becomes effective within $r\lesssim10$ au \citep{mac07}, the launching radius of the outflow, $5-16$ au, is consistent with the region where the rotation is dominant.
In contrast, the circum-binary disk is formed in the lower-density outer region ($r\sim100$ au), resulting in the circum-binary disk being aligned with the magnetic field rather than the rotation axis. 
This scenario can explain why the outflow launched from the circum-stellar disk of VLA 1623A1 (or A2) is aligned with the rotation of  the envelope rather than that of the circum-binary disk.

The misalignment between the angular momentum and magnetic field could be a reasonable scenario to explain the misalignment of the circum-binary disk.
Further studies of angular momentum of the circum-stellar disks of VLA 1623A1 and A2 will allow us to further constrain the detailed structures of the VLA 1623A binary system.
The large-scale magnetic field should also be investigated to reveal better the magnetic field direction in the envelope.

\subsection{The origin of the {opposite velocity gradient} of the circumstellar disk around VLA 1623B}\label{subsec:mis_diskb}

As discussed in the previous subsection, the  VLA 1623B disk rotation has the opposite direction to the envelope, outflow, and circum-binary disk.
The opposite rotations or misalignments of the angular momentum have been reported in several protostellar disks such as BHR71 (with a separation of  $\sim3000$ au: \citealt{tob19}), IRAS 16293-2422 (with a separation of $\sim600$ au: \citealt{zap13}), and IRAS 04191+1523 (with a separation of $\sim860$ au: \citealt{leej17}).
These misaligned binary systems are suggested to be caused by turbulent fragmentation of their parent cores \citep[e.g.,][]{pad02,goo04,off10} rather than fragmentation with rotating cores \citep[e.g.,][]{pri07,bos14}.

One difficulty with the turbulent fragmentation for VLA 1623 is the separation of between VLA 1623A and B.
In general, initial separations of the fragments are suggested to be a few 1000s au \citep{off10,tsu13}.
In contrast, the projected distance of the VLA  1623A and B is $\sim150$ au, which is much closer than the other misaligned binary systems noted. 
A separation of 150 au is similar to the case of  the triplet system of L1448 IRS3B, which is thought to be  formed by disk fragmentation via the gravitational instability  \citep{tob16}. 
However, it might be possible that the separation of the VLA  1623A and B is much larger than 150 au. 
We found that VLA 1623B is located behind VLA 1623A, because the outflow emission is masked by the optically thick background emission of the VLA  1623B disk.
The lack of any indication of gravitational interaction between the circum-binary disk and the VLA 1623 B disk (e.g., through disturbances in the density or velocity structure at the edges of those disks) is also notable, and suggests an association in projection only.
Therefore, it should be noted that the separation of 150 au is a lower limit.

Another difficulty is that turbulent motions were not identified in the envelope.
The typical linewidth of the H$^{13}$CO$^+$ is $\Delta v\sim0.3-0.5$ km s$^{-1}$, and an ordered rotation pattern is clearly observed in the envelope.
These observed features indicate that the turbulence is not significant in this region in comparison with other regions such as BHR71 \citep{tob19}.

Non-ideal MHD simulations show that the counter-rotation can appear in the outflow on envelope scales due to the Hall effect \citep[e.g.,][]{kra11,tsu15,wru16,zhao21}. However, it is not clear whether such counter-rotation can form a protostar and an accretion disk.

One possibility is that the mutual gravito-hydrodynamical interaction between the circum-binary disk and circumstellar disk disturbs the orbit of the tertiary disk, resulting in the reverse rotation \citep{tak21}.
\citet{tak21} simulated the evolution of a triplet protostellar system by using three-dimensional smoothed particle hydrodynamics (SPH) without magnetic fields.
They found that the tertiary protostar is formed via the circum-binary disk fragmentation and the initial rotational directions of all the three circumstellar disks are almost parallel to that of the orbital motion of the binary system. 
Then, the tertiary orbit becomes unstable due to the three-body effect, and  the circum-stellar disk of the tertiary orbit receives the counter-rotating gas accretion.
As a result, the rotational direction becomes reverse.
This scenario seems to be consistent with our results because the close encounter or three-body interaction was suggested \citep{mur13,har18}.

 An interaction with the outflow might also change the rotation direction of the VLA 1623B disk.
The VLA 1623B disk has previously been interpreted as a knot created by outflow shock \citep{bon97,mau12,har20} because VLA 1623B (and also source W) is in the outflow cavity.
Furthermore, the CO brightness temperature is $\sim90$ K at the position of VLA 1623B.
The morphology and the local temperature enhancement  indicate the gas is heated, possibly due to a shock in the outflow.
Therefore, the VLA 1623B protostar may be affected by the outflow launched from VLA 1623A1.
The interaction with the outflow might have changed the disk direction even if the disk were originally aligned with the envelope and circum-binary disk.
However, more detailed studies of the possible time evolution, through the proper motions and the temporal changes in the inclination of the disks, for example, are needed to confirm this scenario.

Another possible scenario might be a collision of the accretion flows from the envelope.
\citet{hsi20} identified several accretion flows connecting to VLA 1623B.
They suggested that the blue- and red-shifted accretion flows collide with each other at  VLA 1623B, which may remove and/or change the angular momentum of the accretion.
In this case, a disk with the opposite rotation might be formed.
We need further molecular line observations with high-sensitivity and high spatial resolution to investigate the accretion flows and their  accretion directions in detail.

\section{summary}\label{sec:sum}

As part of the ALMA Large Program FAUST, we have investigated the kinematics of the triple protostellar system of the VLA 1623$-$2427 low-mass star forming region using H$^{13}$CO$^+$ ($J=3-2$), CS ($J=5-4$), and CCH ($J=7/2-5/2$) emission. 
We have identified the rotation motions on large scales of the envelope and outflows, and on a smaller scale (50 au), of the circum-binary and circumstellar disks.  The main results are listed below.

\begin{itemize}

\item The H$^{13}$CO$^+$ emission traces the envelope and circum-binary disk structures, while the CS and CCH emission mainly trace a wide and low-velocity outflow cavity structure. 
In the mean velocity map of the H$^{13}$CO$^+$ emission we find a twisted pattern, which indicates misalignment by $\sim12\pm6^\circ$ between the rotation axis of the envelope and that of the circum-binary disk: the rotation axis of the envelope is $\sim127^\circ$, while that of the circum-binary disk is $\sim115^\circ$.  The outflow direction is $\sim125^\circ$ and  $\sim129^\circ$ (from CS and CCH), in agreement with the envelope and not with the circum-binary disk.

 \item We find that the velocity gradient of the outflow cavity structure is traced by CS and CCH emission, which can be explained either by rotation of the low-velocity outflow or  entrainment of the two high-velocity outflows. If we assume the case of the outflow rotation, the launching radius of the low-velocity outflow is estimated to be $r\sim5-16$ au from the protostar.
This suggests that the low-velocity outflow is launched from the outer edge of the VLA 1623A1 disk by accreting materials to the disk from the circum-binary disk, while the high-velocity outflows are launched from the inner edges of the circum-stellar disks.
Future observations with higher resolution are needed to assess this scenario.

\item The magnetic field appears to be parallel to the rotation axis of the circum-binary disk, supporting the idea that the circum-binary disk was formed in a low-density outer region coupled with the magnetic field. In contrast, the circum-stellar disk is formed in a high-density inner region. Therefore, the misalignment between the envelope rotation and magnetic field direction could cause the misalignment of the circum-binary disk. This picture is consistent with the non-ideal MHD simulations studied by \citet{hir19}.

\item We also find that the high-velocity CS emission traces the disk rotation around the VLA 1623B protostar. The velocity gradient of the VLA 1623B disk is opposite to those of the other components of the region. The dynamical mass of the VLA 1623B is suggested to be higher than the total mass of the VLA 1623A protostars. Although detailed origins of such opposite velocity gradient is not clear and calls for further investigations, the hydrodynamical interaction with the circum-binary disk or the interaction with the outflow are possible causes of the change in the disk direction.

\end{itemize}


We gratefully appreciate the comments from the anonymous referee that significantly improved this article.
S.O. thanks Masahiro N. Machida, Yu Saiki, Hauyu Baobab Liu for fruitful discussions.
This paper makes use of the following ALMA data set:
ADS/JAO.ALMA\# 2018.1.01205.L (PI: Satoshi Yamamoto). ALMA is a partnership of the ESO (representing its member states), the NSF (USA) and NINS (Japan), together with the NRC (Canada) and the NSC and ASIAA (Taiwan), in cooperation with the Republic of Chile.
The Joint ALMA Observatory is operated by the ESO, the AUI/NRAO, and the NAOJ. The National Radio Astronomy Observatory is a facility of the National Science Foundation operated under cooperative agreement by Associated Universities, Inc. The authors thank the ALMA staff for their excellent support.

This work is supported by the projects PRIN-INAF 2019 ``Planetary systems at young ages (PLATEA)'' and the European Research Council (ERC) under the European Union's Horizon 2020 research and innovation program, for the Project “The Dawn of Organic Chemistry” (DOC), grant agreement No 741002; the PRIN-INAF  The Cradle of Life - GENESIS-SKA (General Conditions in Early Planetary Systems for the rise of life with SKA);  the European Union’s Horizon 2020 research and innovation programs under projects “Astro-Chemistry Origins” (ACO), Grant No 811312.

This project is also supported by a Grant-in-Aid from Japan Society for the Promotion of Science (KAKENHI: Nos. 18H05222, 19H05069, 19K14753, 20K14533, 20H05844, 20H05845).

I.J.-S. has received partial support from the Spanish State Research Agency (AEI; project number PID2019-105552RB-C41).

Data analysis was in part carried out on common use data analysis computer system at the Astronomy Data Center, ADC, of the National Astronomical Observatory of Japan.



{}

\end{document}